\documentclass[11pt,epsf]{article}

\usepackage{graphicx}
\usepackage{amsmath}
\usepackage{color}
\usepackage{cite}
\usepackage{verbatim}
\definecolor{darkgray}{gray}{.4}
\newcommand{\DG}[3][]{{\color{blue}**#2**} {\color{darkgray}[REPLACES [#3] BECAUSE [#1]]}}

\newcommand{\intd}{{\,\normalfont{\text d}}} 
\newcommand{\Hb}{\mathcal{H}_2}

\newcommand{\eqd}{\stackrel{\triangle}{=}}
\newcommand {\exe} {\stackrel{\cdot} {=}}
\newcommand {\lexe} {\stackrel{\cdot} {\le}}

\newcommand {\reals} {{\rm I\!R}}

\newcommand {\bh} {\mbox{\boldmath $h$}}

\newcommand {\bmu} {\mbox{\boldmath $\mu$}}

\newcommand {\bp} {\mbox{\boldmath $p$}}

\newcommand {\br} {\mbox{\boldmath $r$}}
\newcommand {\bs} {\mbox{\boldmath $s$}}

\newcommand {\bu} {\mbox{\boldmath $u$}}
\newcommand {\bv} {\mbox{\boldmath $v$}}

\newcommand {\bx} {\mbox{\boldmath $x$}}
\newcommand {\by} {\mbox{\boldmath $y$}}

\newcommand {\bE} {\mbox{\boldmath $E$}}

\newcommand {\bH} {\mbox{\boldmath $H$}}

\newcommand {\bN} {\mbox{\boldmath $N$}}

\newcommand {\bS} {\mbox{\boldmath $S$}}

\newcommand {\bW} {\mbox{\boldmath $W$}}
\newcommand {\bX} {\mbox{\boldmath $X$}}
\newcommand {\bY} {\mbox{\boldmath $Y$}}

\newcommand{\calC}{{\cal C}}

\newcommand{\calE}{{\cal E}}

\newcommand{\calN}{{\cal N}}

\newcommand{\calX}{{\cal X}}

\begin{document}
\thispagestyle{empty}
\setcounter{page}{1}
\title{Statistical Physics of Signal Estimation in 
  Gaussian Noise: Theory and Examples of Phase Transitions\thanks{The
    work of D.~Guo is supported by the NSF under grant CCF-0644344 and
    DARPA under grant W911NF-07-1-0028 .  The work of S.~Shamai is
    supported in part by the Israel Science Foundation.}}
\author{Neri Merhav,\thanks{N.~Merhav is with the Department of Electrical
    Engineering, Technion -- Israel Institute of Technology, Haifa
    32000, Israel. E--mail: {\tt merhav@ee.technion.ac.il}} Dongning
  Guo,\thanks{D.~Guo is with the Department of Electrical Engineering
    and Computer Science,
Northwestern University,
Evanston, IL 60208, U.S.A. E--mail: {\tt dguo@northwestern.edu}} 
and Shlomo Shamai (Shitz)\thanks{S.~Shamai 
is with the Department of Electrical 
Engineering, Technion -- Israel Institute 
of Technology, Haifa 32000, Israel. E--mail: {\tt sshlomo@ee.technion.ac.il}}
} 
\maketitle

\vspace{-.5in}

\begin{abstract}
We consider the problem of signal estimation (denoising) from a statistical
mechanical perspective, using a
relationship between the minimum mean square error (MMSE), of estimating a
signal, and the mutual information between this signal and its noisy version. The
paper consists of essentially two parts. In the first,
we derive several statistical--mechanical relationships between a 
few important quantities in this
problem area, such as the MMSE, the differential entropy, the Fisher information, 
the free energy, and a
generalized notion of temperature. We also draw analogies
and differences between certain relations pertaining to the
estimation problem and the parallel relations in
thermodynamics and statistical physics. In the second part of the paper, we
provide several application examples, where we demonstrate how certain
analysis tools that are customary in statistical physics,
prove useful in the analysis of the MMSE. In most of these examples, the
corresponding statistical--mechanical systems turn out to consist of
strong interactions that cause phase transitions, which in turn are reflected
as irregularities and discontinuities 
(similar to threshold effects) in the behavior of the MMSE.
\end{abstract}


\section{Introduction}

The relationships and the interplay between Information Theory
and Statistical Physics have been recognized and exploited for several decades
by now. The roots of these relationships date back to the 
celebrated papers by Jaynes from the late fifties of
the previous century \cite{Jaynes57a, Jaynes57b}, but their aspects and
scope have been vastly expanded and deepened ever since. Much of the research
activity in this interdisciplinary problem area revolves around the
identification of `mappings' between problems in Information Theory
and certain many--particle systems in Statistical Physics, which are analogous
at least as far as their mathematical formalisms go. One important example is
the paralellism and analogy between random code ensembles in Information Theory and certain
models of disordered magnetic materials, known as {\it spin glasses}. This
analogy was first identified by Sourlas (see, e.g.,
\cite{Sourlas89, Sourlas94}) and has been further studied in the
last two decades to a great extent. Beyond the fact that
these paralellisms and analogies are academically interesting 
in their own right, 
they also prove useful and beneficial. Their utility stems from the fact that
physical insights, as well as
statistical mechanical tools and analysis techniques
can be harnessed in order to advance the knowledge and the
understanding with regard to
the information--theoretic problem under discussion.

In this context, our work takes place at the meeting point of
Information Theory, Statistical Physics, and yet another area -- Estimation
Theory, where the bridge between information--theoretic and the
estimation--theoretic ingredients of the topic under discussion
is established by an identity \cite[Theorem 2]{GSV05}, 
equivalent to the de Bruijn identity (cf.\ e.g., \cite[Theorem 17.7.2]{CT06}), which 
relates the minimum mean square error (MMSE), of estimating a
signal in additive white Gaussian noise (AWGN),
to the mutual information between this signal and its noisy version. 
We henceforth refer to this relation as the {\it I--MMSE relation}.
It should be pointed out that the present work is not the first to deal with
the interplay between
the I--MMSE relation and statistical mechanics. In an earlier paper
by Shental and Kanter \cite{SK08}, the main theme was an attempt to provide
an alternative proof of the I--MMSE relation, which is rooted in thermodynamics and
statistical physics. However, to this end, the authors of \cite{SK08} had to
generalize the theory of thermodynamics. 

Our study is greatly triggered by
\cite{SK08} (in its earlier versions), but it takes a substantially different route.
Rather than proving the I--MMSE relation,
we simply use it in conjunction
with analysis techniques used in
statistical physics. The basic 
idea that is underlying our work is that 
when the channel input signal is rather complicated 
(but yet, not too complicated), 
which is the case in certain
applications, the mutual information with its
noisy version can be evaluated using statistical--mechanical analysis
techniques, and then related to the MMSE using the I--MMSE relation. This
combination proves rather powerful, because it enables one to distinguish
between situations where irregular (i.e., non--smooth or even
discontinuous) behavior of the mean square error
(as a function of the signal--to--noise ratio) is due to artifacts of
a certain ad--hoc signal estimator, and situations where these irregularities
are inherent in the model, in the sense that they are apparent even 
in optimum estimation. In the latter situations,
these irregularities (or 
threshold effects) are intimately related to {\it phase transitions} in the
parallel statistical--mechanical systems.

These motivations set the stage for our study of the relationships between the
MMSE and statistical mechanics, first of all, in the general level, and then
in certain concrete applications.
Accordingly, the paper consists of two main parts. In the first,
which is a general theoretical study,
we derive several statistical--mechanical relationships between a
few important quantities 
such as the MMSE, the differential entropy, the Fisher
information,
the free energy, and a
generalized notion of temperature. We also draw analogies
and differences between certain relations pertaining to the
estimation problem and the parallel relations in
thermodynamics and statistical physics. In the second part of the paper, we
provide several application examples, where we demonstrate how certain
analysis tools that are customary in statistical physics (in conjunction with
large deviations theory)
prove useful in the analysis of the MMSE. 
In light of the motivations described in the previous paragraph, in most of these examples, the
corresponding statistical--mechanical systems turn out to consist of
strong interactions that cause phase transitions, which in turn are reflected
as irregularities and discontinuities
in the behavior of the MMSE.

The remaining part of this paper is organized as follows:
In Section \ref{s:notation},
we establish a few notation conventions and we formalize the
setting under discussion. In Section \ref{physbg}, 
we provide the basic background in
statistical physics that will be used in the sequel. Section \ref{s:th}
is devoted to
the general theoretical study, and finally, Section \ref{s:ex}
includes application
examples, where the MMSE will be analyzed using statistical--mechanical tools.

\section{Notation Conventions, Formalization and Preliminaries}
\label{s:notation}
\subsection{Notation Conventions}

Throughout this paper, scalar random
variables (RV's) will be denoted by capital
letters, like $X$ and $Y$, their sample values will be denoted by
the respective lower case letters, and their alphabets will be denoted
by the respective calligraphic letters.
A similar convention will apply to
random vectors and their sample values,
which will be denoted with the same symbols in the boldface font.
Thus, for example, $\bX$ will denote a random $n$-vector $(X_1,\ldots,X_n)$,
and $\bx=(x_1,...,x_n)$ is a specific vector value in $\calX^n$,
the $n$-th Cartesian power of $\calX$.

Sources and channels will be denoted generically by the letters $P$ and $Q$.
The expectation operator will be
denoted by $\bE\{\cdot\}$. When the underlying probability measure
is indexed by a parameter, say, $\beta$, then it will used as a subscript
of $P$, $p$ and $\bE$, unless there is no ambiguity.

For two positive sequences $\{a_n\}$ and $\{b_n\}$, the notation
$a_n\exe b_n$ means that $a_n$ and $b_n$ are asymptotically of the same
exponential order, that is, $\lim_{n\to\infty}\frac{1}{n}\ln\frac{a_n}{b_n}
=0$. Similarly, $a_n\lexe b_n$ means that
$\limsup_{n\to\infty}\frac{1}{n}\ln\frac{a_n}{b_n}\le 0$,
etc. Information theoretic quantities like entropies and mutual
informations will be denoted following the usual conventions
of the Information Theory literature.

\subsection{Formalization and Preliminaries}

We consider the simplest variant of the signal 
estimation problem setting studied in \cite{GSV05},
with a few slight modifications in notation. 
Let $(\bX,\bY)$ be a pair of random vectors in $\reals^n$, related by the Gaussian channel
\begin{equation}
\label{GCv}
\bY=\bX+\bN,
\end{equation}
where $\bN$ is a random vector (noise), whose components are i.i.d.,
zero--mean, Gaussian random variables (RV's) whose variance is
$1/\beta$, where $\beta$ is a given positive constant designating the
signal--to--noise ratio (SNR), or the inverse temperature in
statistical--mechanical point of view (cf.\ Section \ref{physbg}).
It is assumed that $\bX$ and $\bN$ are independent. Upon receiving $\bY$,
one is interested in inferring about the (desired) random vector $\bX$. As is well known, the
best estimator of $\bX$ given the observation vector $\bY$, in the mean square error (MSE)
sense, i.e., the MMSE estimator, is the
conditional mean $\hat{\bX}=\bE(\bX|\bY)$ and the corresponding MMSE,
$\bE\|\hat{\bX}-\bX\|^2$ will
denoted by $\mbox{mmse}(\bX|\bY)$. Theorem 2 in \cite{GSV05},
which provides the I--MMSE relation, relates the
MMSE to the mutual information $I(\bX;\bY)$ (defined using the natural base logarithm) 
according to 
\begin{equation}
\label{gsvvector}
\frac{\mbox{d}I(\bX;\bY)}{\mbox{d}\beta}=\frac{\mbox{mmse}(\bX|\bY)}{2}.
\end{equation}
For example, if $n=1$ and $X\sim\calN(0,1)$, then $I(X;Y)=\frac{1}{2}\ln(1+\beta)$,
which leads to
$\mbox{mmse}(X|Y)=1/(1+\beta)$, in agreement with elementary results.
The relationship has been used in \cite{PalVer07IT} to compute the
mutual information achieved by low-density parity-check (LDPC) codes
over Gaussian channels through evaluation of the marginal estimation
error.

A very important function, which will be pivotal to our derivation of
both $\bE(\bX|\bY)$ and $\mbox{mmse}(\bX|\bY)$, as well as to the
mutual information $I(\bX;\bY)$, is the posterior distribution.
Denoting the probability 
mass 
function of $\bx$ by $Q(\bx)$ and the
channel induced by (\ref{GCv}) by $P(\by|\bx)$, then 
\begin{align}
\label{posterior}
P(\bx|\by)&=\frac{Q(\bx)P(\by|\bx)}{\sum_{\bx'}Q(\bx')P(\by|\bx')}\nonumber\\
&=
\frac{Q(\bx)\exp[-\beta\cdot\|\by-\bx\|^2/2]}{Z(\beta|\by)},
\end{align}
where we defined 
\begin{equation}
\label{partitionfunction}
Z(\beta|\by)\eqd\sum_{\bx}Q(\bx)\exp[-\beta\cdot\|\by-\bx\|^2/2]
= (2\pi/\beta)^{n/2} P_\beta(\by)
\end{equation}
where $P_\beta(\by)$ is the channel output density.
Here we have assumed that $\bx$ is discrete, as otherwise $Q$
should be replaced by the 
probability density function (pdf) and the summation
over $\{\bx'\}$ should be replaced by an integral. 
The function $Z(\beta|\by)$ is very similar to the so-called
{\it partition function}, which is
well known to play a very central role in statistical mechanics, and
will also play a central role in our analysis. In the next section,
we then give some necessary background in statistical mechanics that
will be essential to our study. 

\section{Physics Background}
\label{physbg}

Consider a physical system with $n$ particles,
which can be in a variety of microscopic states (`microstates'),
defined by combinations of physical quantities associated with
these particles, e.g.,
positions, momenta,
angular momenta, spins, etc., of all $n$ particles.
For each such
microstate of the system, which we shall
designate by a vector $\bx=(x_1,\ldots,x_n)$, there is an
associated energy, given by a {\it Hamiltonian} (energy function),
$\calE(\bx)$. For example, if $x_i=(\bp_i,\br_i)$, where
$\bp_i$ is the momentum vector of particle number $i$ and
$\br_i$ is its position vector, then classically, $\calE(\bx)=\sum_{i=1}^N
\left[\frac{\|\bp_i\|^2}{2m}+mgz_i\right]$, where $m$ is the mass of each particle,
$z_i$ is its height -- one of the
coordinates of $\br_i$, and
$g$ is the gravitation constant.

One of the most
fundamental results in statistical physics (based on the
law of energy conservation and
the basic postulate that all microstates of
the same energy level are equiprobable)
is that when the system is in thermal
equilibrium with its environment, the probability of finding the system in 
a microstate $\bx$ is
given by the {\it Boltzmann--Gibbs} distribution
\begin{equation}
\label{bd}
P(\bx)=\frac{e^{-\beta\calE(\bx)}}{Z(\beta)}
\end{equation}
where $\beta=1/(kT)$, $k$ being Boltmann's constant and $T$ being temperature,
and
$Z(\beta)$ is the normalization constant,
called the {\it partition function}, which
is given by
$$Z(\beta)=\sum_{\bx} e^{-\beta\calE(\bx)},$$
assuming discrete states.  In case of continuous state space, the partition
function is defined as
$$Z(\beta)=\int \mbox{d}\bx\, e^{-\beta\calE(\bx)},$$
and $P(\bx)$ is understood as a pdf.
 The role
of the partition function is by far deeper
than just being a normalization factor, as
it is actually the key quantity from which many
macroscopic physical quantities can be derived,
for example, the free energy\footnote{The free
energy means the maximum work that the system
can carry out in any process of fixed temperature.
The maximum is obtained when the process is reversible (slow, quasi--static
changes in the system).}
is $F(\beta)=-\frac{1}{\beta}\ln Z(\beta)$, the average internal
energy
is given by $\bar{E}\eqd\bE\{\calE(\bX)\}=-(\mbox{d}/\mbox{d}\beta)\ln Z(\beta)$ with $\bX \sim P(\bx)$, the heat capacity
is obtained from
the second derivative, etc. One of the ways to
obtain eq.\ (\ref{bd}), is as the maximum entropy
distribution under an average energy constraint
(owing to the second law of thermodynamics), where $\beta$ plays the role of
a Lagrange multiplier that controls the average energy.

An important special case, which is very relevant both in physics 
and in the study of AWGN channel considered here, is the case where the
Hamiltonian $\calE(\bx)$ is additive and {\it quadratic} (or ``harmonic'' in
the physics terminology), i.e.,
$\calE(x)=\sum_{i=1}^n\frac{1}{2}\kappa x_i^2$, 
for some constant $\kappa > 0$, or even more generally,
$\calE(x)=\sum_{i=1}^n\frac{1}{2}\kappa_i x_i^2$, which means that the components
$\{x_i\}$ are Gaussian and independent. A classical result in this case, 
known as the {\it equipartition theorem of energy}, which is 
very easy to show, asserts that each particle (or, more 
precisely, each degree of freedom) contributes an average energy
of $\bE\{\frac{1}{2}\kappa_i X_i^2\}=1/(2\beta)=kT/2$ independently of
$\kappa$ (or $\kappa_i$).

Returning to the case of a general Hamiltonian, it is instructive to
relate the Shannon entropy, pertaining to the Boltzmann--Gibbs distribution,
to the quantities we have seen thus far. Specifically,
the Shannon entropy $S(\beta)=-\bE\{\ln P(\bX)\}$ associated with
$P(\bx)=e^{-\beta\calE(\bx)}/Z(\beta)$, 
is given by
$$S(\beta)=
\bE\ln\left[\frac{Z(\beta)}{e^{-\beta\calE(\bx)}}\right]
=\ln Z(\beta)+\beta\cdot\bar{E},$$
where, as mentioned above,
\begin{equation}
\label{energy}
\bar{E}=-\frac{\mbox{d}\ln Z(\beta)}{\mbox{d}\beta}
\end{equation}
is the average internal energy. This suggests
the differential equation
\begin{equation}
\label{diffeq}
\dot{\psi}(\beta)-\frac{\psi(\beta)}{\beta}=\frac{S(\beta)}{\beta},
\end{equation}
where $\psi(\beta)=-\ln Z(\beta)$ and $\dot{\psi}$ means the derivative of
$\psi$. Equivalently, eq.\ (\ref{diffeq}) can be rewritten as:
\begin{equation}
\beta\frac{\mbox{d}}{\mbox{d}\beta}\left[\frac{\psi(\beta)}{\beta}\right]=\frac{S(\beta)}{\beta},
\end{equation}
whose solution is easily found to be
\begin{equation}
\label{psisolution}
\psi(\beta)=\beta E_0-
\beta\int_\beta^\infty \frac{d\hat{\beta}S(\hat{\beta})}{\hat{\beta}^2},
\end{equation}
where $E_0=\min_{\bx}\calE(\bx)$ is the {\it ground--state energy}, here
obtained as a constant
of integration by examining the limit of $\beta\to\infty$.
Thus, we see that the log--partition function at a given temperature can be expressed as a
{\it heat integral} of the entropy, namely, as an integral
of a function that consists of the entropy at all lower temperatures.
This is different from the other relations we mentioned thus far, which were
all `pointwise' in the temperature domain, in the sense that all quantities were
pertaining to the same temperature. Taking the derivative of $\psi(\beta)$
according to eq.\ (\ref{psisolution}), we obtain the average internal energy:
\begin{equation}
\label{internalenergy}
\bar{E}=\dot{\psi}(\beta)=E_0-\int_\beta^\infty
\frac{d\hat{\beta}S(\hat{\beta})}{\hat{\beta}^2}+\frac{S(\beta)}{\beta},
\end{equation}
where the first two terms form the free energy.\footnote{By changing the
integration variable from $\beta$ to $T$, this is identified with the relation
$F=E_0-\int_0^T S\mbox{d}T'$, which together with $F=\bar{E}-ST$, complies with
the relation $\bar{E}=E_0+\int_0^ST\mbox{d}S'=E_0+\int_0^Q\mbox{d}Q'$, 
accounting for the simple fact that in the absence of any external work
applied to the system,
the internal energy is simply the heat accumulated as temperature is raised
from $0$ to $T$.}

As a final remark,
we should note that although the expression $Z(\beta|\by)$ of 
eq.\ (\ref{partitionfunction}) is similar to that of $Z(\beta)$ defined in
this section (for a quadratic Hamiltonian), there is nevertheless a small
difference: The exponentials in (\ref{partitionfunction}) are weighted
by probabilities $\{Q(\bx)\}$, which are independent of $\beta$.
However, as explained in \cite[p.\ 3713]{Merhav08}, this is not an essential difference
because these weights can be interpreted as degeneracy of states, that is, as
multiple states (whose number is proportional to $Q(\bx)$) of the same energy.

\section{Theoretical Derivations}
\label{s:th}

Consider the Gaussian channel (\ref{GCv}) and the corresponding posterior
(\ref{posterior}).
Denoting by $\bE_\beta$ the expectation operator w.r.t.\ joint pdf
of $(\bX,\bY)$ induced by $\beta$, we have:
\begin{align}
\label{start}
I(\bX;\bY)
&=\bE_\beta\left\{\ln\frac
{\exp[-\beta\cdot\|\bY-\bX\|^2/2]}
{Z(\beta|\bY)}
\right\}
\nonumber\\
&=
- \frac{\beta}{2}\,\bE_\beta\left\{\|\bY-\bX\|^2\right\}
- \bE_\beta\left\{\ln Z(\beta|\bY)\right\}
\nonumber\\
&=
-\frac{n}{2} -
\bE_\beta\left\{\ln Z(\beta|\bY)\right\}
\end{align}
where we use the fact that $\bE_\beta\left\{\|\bY-\bX\|^2\right\}
= \bE_\beta\left\{ \|\bN\|^2 \right\} = n/\beta$.
Taking derivatives w.r.t.~$\beta$, and using the I--MMSE relation, we then
have:
\begin{equation}
\frac{\mbox{mmse}(\bX|\bY)}{2}=\frac{\partial I(\bX;\bY)}{\partial\beta}
=-\frac{\partial}{\partial\beta}\bE_\beta\{\ln Z(\beta|\bY)\}.
\end{equation}
and so, we obtain a very simple relation between the MMSE and the partition function
of the posterior:
\begin{equation}
\label{main}
\mbox{mmse}(\bX|\bY)=-2\frac{\partial}{\partial\beta}\bE_\beta\{\ln Z(\beta|\bY)\}
\end{equation}
By calculating the derivative of 
the right-hand side (r.h.s.) more explicitly, one further obtains
the following:
\begin{align}
\label{covariance.eq}
-\frac{\partial}{\partial\beta}\bE_\beta\ln Z(\beta|\bY)&=
-\frac{\partial}{\partial\beta}\int_{\reals^n} d\by\cdot 
P_\beta(\by)\ln Z(\beta|\by)\nonumber\\
&=-\int_{\reals^n} d\by\cdot P_\beta(\by)
\frac{\partial\ln Z(\beta|\by)}{\partial\beta}
-\int_{\reals^n}d\by
\cdot\frac{\partial P_\beta(\by)}{\partial\beta}\cdot\ln Z(\beta|\by).
\end{align}
Now, the first term at the right--most side of (\ref{covariance.eq}) 
can easily be computed 
by using the fact that $\ln Z(\beta|\by)$ is a log--moment generating
function of the energy 
(as is customarily done in statistical mechanics,
cf.\ eq.\ (\ref{energy})), which implies that
it is given by 
$\bE_\beta\{\|\bY-\bX\|^2\}=n/(2\beta)=nkT/2$, just like in the energy
equipartition theorem for quadratic Hamiltonians. As for the second 
term, we have
\begin{align}
& \int_{\reals^n}d\by
\cdot\frac{\partial P_\beta(\by)}{\partial\beta}\cdot\ln
Z(\beta|\by)\nonumber\\
&=
\int_{\reals^n}d\by\cdot P_\beta(\by)
\cdot\frac{\partial \ln P_\beta(\by)}{\partial\beta}\cdot\ln Z(\beta|\by)
\nonumber\\
&=\int_{\reals^n}\mbox{d}\by\cdot\left(\frac{2\pi}{\beta}\right)^{-n/2} 
\sum_{\bx}Q(\bx)\left[\frac{n}{2\beta}-\frac{1}{2}\|\by-\bx\|^2\right]\cdot
\exp\{-\beta\|\by-\bx\|^2/2\}\ln Z(\beta|\by)\nonumber\\
&=-\frac{1}{2}
\mbox{Cov}\{\|\bY-\bX\|^2,\ln Z(\beta|\bY)\}.
\end{align}
The MMSE is then given by
\begin{equation}
\mbox{mmse}(\bX|\bY)=-2\frac{\partial}{\partial\beta}\bE_\beta\{\ln Z(\beta|\bY)\}
=\frac{n}{\beta}+\mbox{Cov}\{\|\bY-\bX\|^2,\ln Z(\beta|\bY)\},
\end{equation}
which can then be viewed as a variant of the energy equipartition theorem
with a correction term that stems from the fact the pdf of $\bY$
depends on $\beta$.

Another look, from an estimation--theoretic point of view,
at this expression reveals the following:
The first term, $n/\beta=\bE\|\bY-\bX\|^2$, is the amount of 
noise in the raw data $\bY$, without any processing.
The second term, which is always negative, designates then the {\it noise suppression level}
due to MMSE estimation relative to the raw data. 
The intuition behind the covariance term is that when the
`correct' $\bx$ (the one that actually feeds the Gaussian channel) 
dominates the partition function then 
$\ln Z(\beta|\bY)\approx-\beta\|\bY-\bX\|^2/2$, and so,
there is a very strong negative correlation between 
$\|\bY-\bX\|^2$ and $\ln Z(\beta|\bY)$. In particular,
\begin{equation}
\mbox{Cov}\{\|\bY-\bX\|^2,-\beta\|\bY-\bX\|^2/2\}=-\frac{n}{\beta},
\end{equation}
which exactly cancels the above--mentioned first term, $n/\beta$, and so, the
overall MMSE essentially vanishes. When the correct $\bx$ is not dominant,
this correlation is weaker.
Also, note that since
\begin{equation}
\bE\|\bY-\bX\|^2=\mbox{mmse}(\bX|\bY)+\bE\|\bY-\bE(\bX|\bY)\|^2,
\end{equation}
then this implies that
\begin{equation}
\bE\|\bY-\bE(\bX|\bY)\|^2=-\mbox{Cov}\{\|\bY-\bX\|^2,\ln Z(\beta|\bY)\}.
\end{equation}
It is now interesting to relate the noise suppression level
$$\Delta\eqd \bE\|\bY-\bE(\bX|\bY)\|^2=-\mbox{Cov}\{\|\bY-\bX\|^2,\ln Z(\beta|\bY)\}$$
to the Fisher information matrix and then to a new generalized notion of temperature due to Narayanan
and Srinivasa \cite{NS07} via the de Bruijn identity. According to de Bruijn's identity,
if $\bW$ is a vector of i.i.d.\ standard normal components, independent of $\bX$, then
$$\frac{\mbox{d}}{\mbox{d}t} h(\bX+\sqrt{t}\bW)=\frac{1}{2}\mbox{tr}\{J(\bX+\sqrt{t}\bW)\}$$
where $h(\bY)$ is differential entropy and
$J(\bY)$ is the Fisher information matrix associated with $\bY$ w.r.t.\ a translation
parameter, namely,
$$\mbox{tr}\{J(\bY)\}=\sum_{i=1}^n\bE\left\{\left[\frac{\partial\ln
P_\beta(\by)}{\partial y_i}\bigg|_{\by=\bY}\right]^2\right\}
=\sum_{i=1}^n\int_{\reals^n}\frac{\mbox{d}\by}{P_\beta(\by)}
\left[\frac{\partial P_\beta(\by)}{\partial y_i}\right]^2.$$
Note that since $P_\beta(\by)$ and $Z(\beta|\by)$ differ only by a multiplicative factor of
$(\beta/2\pi)^{n/2}$, it is obvious that 
$\partial\ln P_\beta(\by)/\partial y_i=
\partial\ln Z(\beta|\by)/\partial y_i$ and so, the Fisher information can also be related directly to
the free energy by
\begin{align}
\mbox{tr}\{J(\bY)\}&=
\sum_{i=1}^n\bE\left\{\left[\frac{\partial\ln Z(\beta|\by)}
{\partial y_i}\bigg|_{\by=\bY}\right]^2\right\}\nonumber\\
&=\sum_{i=1}^n\bE\{[\bE\{-\beta(Y_i-X_i)|\bY\}]^2\}\nonumber\\
&=\beta^2\sum_{i=1}^n\bE\{\bE^2(N_i|\bY)\},
\end{align}
where $N_i=Y_i-X_i$ and where we have used the fact that the
derivative of $\exp\{-\beta\|\by-\bx\|^2\}$ w.r.t.\ $y_i$ is given by
$-\beta(y_i-x_i)\cdot\exp\{-\beta\|\by-\bx\|^2\}$.
Now, as is also shown in \cite{GSV05}:
\begin{align}
I(\bX;\bX+\bN)&=I(\bX;\bX+\bW/\sqrt{\beta})\nonumber\\
&=h(\bX+\bW/\sqrt{\beta})-h(\bW/\sqrt{\beta})\nonumber\\
&=h(\bX+\bW/\sqrt{\beta})-
\frac{n}{2}\ln\left(2\pi e/\beta\right).
\end{align}
Thus,
\begin{align}
\mbox{mmse}(\bX|\bX+\bN)&=2\cdot\frac{\partial I(\bX;\bX+\bN)}{\partial\beta}\nonumber\\
&=2\cdot\frac{\partial h(\bX;\bX+\bW/\sqrt{\beta})}{\partial\beta}+\frac{n}{\beta}\nonumber\\
&=-\frac{1}{\beta^2}\mbox{tr}\{J(\bY)\}+\frac{n}{\beta},
\end{align}
where the factor $-1/\beta^2$ in front of the Fisher information term accounts
for the passage from the variable $t$ to the variable $\beta=1/t$, as $\intd t/\intd \beta=-1/\beta^2$.
Combining this with the previously obtained relations, 
we see that the noise suppression level due to MMSE estimation is given by
$$\Delta=\frac{\mbox{tr}\{J(\bY)\}}{\beta^2}.$$
In \cite[Theorem 3.1]{NS07}, a generalized definition of 
the inverse temperature is proposed, as the response of the entropy to 
small energy perturbations, using de Bruijn's identity. As a consequence of that
definition, the 
generalized inverse temperature 
in \cite{NS07} turns out to be proportional to the Fisher information of $\bY$,
and thus, in our setting, it is also proportional to $\beta^2\Delta$.\footnote{
As is shown in \cite{NS07}, the generalized inverse
temperature coincides with the ordinary inverse temperature when $\bY$ is purely Gaussian
with variance proportional to $1/\beta$, i.e., the ordinary
Boltzmann distribution with a quadratic Hamiltonian. In our setting, 
on the other hand, $\bY$ is given
by a mixture of Gaussians whose weights are independent of $\beta$. To avoid confusion,
it is important to emphasize that the original parameter $\beta$, in our setting, pertains
to the Boltzmann form of the distribution 
of $\bX$ given $\bY=\by$ according to the posterior $P(\bx|\by)$, whereas the current
discussion concerns the 
temperature associated with the (unconditional) ensemble of $\bY=\bX+\bN$.}
It should be pointed out that whenever the system undergoes
a phase transition (as is the case with most of our forthcoming examples), then
$\Delta$, and hence also the effective temperature, may exhibit a 
non--smooth behavior, or even a discontinuity. 

Additional relationships can be obtained in analogy to certain relations
in statistical thermodynamics that were mentioned in Section \ref{physbg}: 
Consider again the chain of equalities (\ref{start}), but this time,
instead using the relation $\bE_\beta\{\|\bY-\bX\|^2\}=n/\beta$, in the
passage from the second to the third line, we use the relation
$\bE_\beta\{\|\bY-\bX\|^2\}=-\bE_\beta\{\frac{\mbox{d}}{\mbox{d}\beta}\ln
Z(\beta|\bY)\}$ in conjunction with the identity (cf.\ eq.\
(\ref{covariance.eq})):
\begin{align}
\bE_\beta\left\{\frac{\mbox{d}\ln Z(\beta|\bY)}{\mbox{d}\beta}\right\}&=
\frac{\mbox{d}\bE_\beta\{\ln Z(\beta|\bY)\}}{\mbox{d}\beta}-
\int_{\reals^n} \mbox{d}\by\frac{\mbox{d}P_\beta(\by)}{\mbox{d}\beta}\cdot
\ln Z(\beta|\by)\nonumber\\
&=\frac{\mbox{d}\bE_\beta\{\ln Z(\beta|\bY)\}}{\mbox{d}\beta}+
\frac{1}{2}\mbox{Cov}\{\|\bY-\bX\|^2,\ln Z(\beta|\bY)\},
\end{align}
to obtain
\begin{equation}
\bE_\beta\{\ln Z(\beta|\bY)\}-\beta\cdot\frac{\mbox{d}}{\mbox{d}\beta}
\bE_\beta\{\ln Z(\beta|\bY)=\frac{\beta}{2}\mbox{Cov}\{\|\bY-\bX\|^2,
\ln Z(\beta|\bY)\}-I(\bX;\bY).
\end{equation}
Thus, redefining the function $\psi(\beta)$ as
\begin{equation}
\psi(\beta)=-\bE_\beta\{\ln Z(\beta|\bY)\},
\end{equation}
we obtain the following differential equation
which is very similar to (\ref{diffeq}):
\begin{equation}
\label{diffeq1}
\dot{\psi}(\beta)-\frac{\psi(\beta)}{\beta}=\frac{\Sigma(\beta)}{\beta}
\end{equation}
where
\begin{equation}
\Sigma(\beta)=
\frac{\beta}{2}\mbox{Cov}\{\|\bY-\bX\|^2,
\ln Z(\beta|\bY)\}-I(\bX;\bY).
\end{equation}
Thus, the solution to this equation is precisely the
same as (\ref{psisolution}), except that $S(\beta)$ is replaced by
$\Sigma(\beta)$ and the ground--state energy $E_0$ is
redefined as
$$E_0=\bE_\beta\{\min_{\bx}\|\bY-\bx\|^2\}.$$
Consequently, $\mbox{mmse}(\bX|\bY)=2\dot{\psi}(\beta)$, where
$$\dot{\psi}(\beta)=
E_0-\int_\beta^\infty\frac{\mbox{d}\hat{\beta}\Sigma(\hat{\beta})}{
\hat{\beta}^2}+\frac{\Sigma(\beta)}{\beta}$$
and one can easily identify the contributions of the 
free energy and the internal
energy (heat), as was done in Section \ref{physbg}. 

To summarize, we see that the I-MMSE relation gives rise essentially 
similar relations
as in statistical thermodynamics except that 
the ``effective entropy'' $\Sigma(\beta)$
includes correction terms that account for the fact that our ensemble
corresponds to a posterior distribution $P(\bx|\by)$ and the fact
that the distribution of $\bY$ depends on $\beta$.

\section{Examples}
\label{s:ex}

In this section, we provide a few examples
where we show how the asymptotic MMSE can be calculated by using the
I--MMSE relation in conjunction with statistical--mechanical techniques for
evaluating the mutual information, or the partition function pertaining to 
the posterior distribution. 

After the first example, of a Gaussian i.i.d.\ channel
input, which is elementary, we turn to explore three examples where the channel
input is a randomly selected 
codebook vector from a certain ensemble of codebooks that comply with a power
constraint $\frac{1}{n}\bE\{\|\bX\|^2\}\le P_x$. There could be various
motivations for 
MMSE estimation when the desired signal is a codeword: One example is that of a
user that, in addition to its desired signal, receives also a relatively
strong interfering
signal, which carries digital information (a codeword) intended to other users, and which comes
from a codebook whose rate exceeds the capacity of this crosstalk channel
between the interferer and our user, so that the user cannot fully decode
this interference. Nonetheless, our user would like to estimate it as
accurately as possible
in order to subtract it and thereby perform interference cancellation. 

In the first example of a code ensemble (Subsection \ref{simplecode}),
we deal with a simple ensemble of block codes, and we demonstrate that the
MMSE exhibits a phase transition at the value of $\beta$ for which the
channel capacity $C(\beta)=\frac{1}{2}\ln(1+\beta P_x)$ agrees with the coding
rate $R$. The second ensemble (Subsection \ref{bc}) consists of an hierarchical
structure which is suitable for the Gaussian broadcast channel. Here, we will
observe two phase transitions, one corresponding to the weak user and one --
to the strong user. The third ensemble (Subsection \ref{tree}) is also hierarchical, but in 
a different way: here the hierarchy corresponds to that of a tree structured code
that works in two (or more) 
segments. In this case, there could be either one
phase transition or two, depending on the coding rates at the two 
segments (see
also \cite{Merhav09}). Our last example is not related to coding applications,
and it is based on a very simple model of sparse signals which is motivated by
compressed sensing applications. Here we show that phase transitions can be present when
the signal components are strongly correlated.

The statistical--mechanical considerations in this section provide
unique insight into the coding and estimation problems, in particular
by examining the typical behavior of the geometry of the free energy.
This is in fact related to the notion of joint typicality for proving
coding theorems, but 
more concrete geometry is seen due to the
special structures of the code ensembles.
In some of the ensuing examples, the mutual information can also be
obtained through existing channel capacity results from information theory.
In the last example pertaining to sparse signals (Subsection
\ref{sparse}), however, we are not aware of any alternative to the
calculation using statistical mechanical techniques.

\subsection{Gaussian I.I.D.\ Input}
\label{Gaussianiid}

Our first example is very simple:
Here, the components of $\bX$ are zero--mean, i.i.d., Gaussian RV's with variance
$P_x$. In this case, we readily obtain
$$Z(\beta|\by)= \frac{\exp\{-\|\by\|^2/[2(P_x+1/\beta)]\}}{(1+\beta P_x)^{n/2}},$$
thus
$$\ln Z(\beta|\by)=-\frac{n}{2}\ln(1+\beta P_x)-\frac{\|\by\|^2}{2(P_x+1/\beta)}.$$
Clearly,
$$\bE_\beta\ln Z(\beta|\bY)=-\frac{n}{2}\ln(1+\beta P_x)-\frac{n}{2}$$
and its negative derivative is $nP_x/[2(1+\beta P_x)]$, which is indeed half of the MMSE.
Here, we have:
$$\Delta=\frac{n}{\beta}-\frac{nP_x}{1+\beta P_x}=\frac{n}{\beta(1+\beta P_x)}$$
and 
$$\mbox{tr}\{J(\bY)\}=n\bE\left[\frac{Y}{P_x+1/\beta}\right]^2=\frac{n\beta}{1+\beta
P_x}$$
and so, the relation $\mbox{tr}\{J(\bY)\}=\beta^2\Delta$ is easily verified. Thus,
the generalized temperature here is $\beta/(1+\beta P_x)$, which is the reciprocal 
of the variance of the Gaussian output.

\subsection{Random Codebook on a Sphere Surface} 
\label{simplecode}

Let $\bX$ assume a uniform distribution over a codebook
$\calC=\{\bx_1,\ldots,\bx_M\}$, $M=e^{nR}$, where each codeword $\bx_i$
is drawn independently under the uniform distribution over the surface of the $n$--dimensional
sphere, which is centered at the origin, and whose radius is $\sqrt{nP_x}$.
The code is capacity achieving (the input becomes essentially i.i.d.\
Gaussian as $n\to\infty$).  In the following we show that the MMSE
vanishes if the code rate $R$ is below channel capacity, but is no
different than that of i.i.d.\ Gaussian input (without code structure)
if $R$ exceeds the capacity.  We note that such a phase transition has
been shown for good binary codes in general in \cite{PelSan07ETT}
using the I-MMSE relationship.

Here, for a given $\by$, we have:
\begin{align}
Z(\beta|\by)&=\sum_{\bx\in\calC}e^{-nR}\exp[-\beta\|\by-\bx\|^2/2]\nonumber\\
&=e^{-nR}\exp[-\beta\|\by-\bx_0\|^2/2]+
\sum_{\bx\in\calC\setminus\{\bx_0\}}e^{-nR}\exp[-\beta\|\by-\bx\|^2/2]\nonumber\\
&\eqd Z_c(\beta|\by)+Z_e(\beta|\by)
\end{align}
where,
without loss of generality,
we assume $\bx_0$ to be the transmitted codeword.
Now, since $\|\by-\bx_0\|^2$ is typically around $n/\beta$, $Z_c(\beta|\by)$
would typically be about $e^{-nR}e^{-\beta\cdot n/(2\beta)}=e^{-n(R+1/2)}$.
As for $Z_e(\beta|\by)$, we have:
$$Z_e(\beta|\by)\exe e^{-nR}\int_{\reals}\mbox{d}\epsilon N(\epsilon)e^{-\beta n\epsilon},$$
where $N(\epsilon)$ is the number of codewords $\{\bx\}$ in $\calC-\{\bx_0\}$ for which
$\|\by-\bx\|^2/2\approx n\epsilon$, namely, between $n\epsilon$ and $n(\epsilon+\mbox{d}\epsilon)$.
Now, given $\by$,  $N(\epsilon)=\sum_{i=1}^M1\{\bx_i:~\|\by-\bx_i\|^2/2\approx n\epsilon\}$
is the sum of $M$ i.i.d.\ Bernoulli RV's and so, its expectation is
\begin{equation}
\overline{N(\epsilon)}=\sum_{i=1}^M\mbox{Pr}\{
\|\by-\bX_i\|^2/2\approx n\epsilon\}= e^{nR}\mbox{Pr}\{\|\by-\bX_1\|^2/2\approx n\epsilon\}.
\end{equation}
Denoting 
$P_y=\frac{1}{n}\sum_{i=1}^ny_i^2$ (typically, $P_y$ is about $P_x+1/\beta$),
the event $\|\by-\bx\|^2/2\approx n\epsilon$ is equivalent to the event
$\langle \bx,\by \rangle\approx[(P_x+P_y)/2-\epsilon]n$ 
or equivalently,
$$\rho(\bx,\by)\eqd \frac{\langle\bx,\by\rangle}{n\sqrt{P_xP_y}}\approx\frac{\frac{1}{2}(P_x+P_y)-\epsilon}
{\sqrt{P_xP_y}}\eqd\frac{P_a-\epsilon}{P_g}$$
where have defined $P_a=(P_x+P_y)/2$ and $P_g=\sqrt{P_xP_y}$ (the arithmetic and the geometric
means between $P_x$ and $P_y$, respectively).
The probability that a randomly chosen vector $\bX$ on the sphere would have
an empirical correlation coefficient $\rho$ with a given vector $\by$ 
(that is, $\bX$ falls within a cone of half angle $\arccos(\rho)$ around $\by$)
is exponentially $\exp[\frac{n}{2}\ln(1-\rho^2)]$. 
For convenience, let us define
\[
\Gamma(\rho) = \frac12 \ln \left( 1- \rho^2 \right)
\]
so that we can write
$$\mbox{Pr}\{\|\by-\bX_1\|^2/2\approx n\epsilon\}\exe
\exp\left\{n \, \Gamma\left(\frac{P_a-
\epsilon}{P_g}\right)\right\}.$$
From this point and onward, our considerations are very similar to those that
have been used in the random energy model (REM) of spin glasses in
statistical mechanics \cite{Derrida80, Derrida80b, Derrida81},
a model of disordered magnetic materials where the energy levels pertaining to
the various configurations of the system $\{\calE(\bx)\}$ are i.i.d.\ RV's.
These considerations have already been applied in the analogous analysis of
random code ensemble performance, where the randomly chosen codewords give
rise to random scores that play the same role as the random energies of the
REM. The reader is referred to
\cite{Sourlas89},\cite{Sourlas94},\cite[Chapters 5,6]{MM06},
and \cite{Merhav08a} for a more detailed account of these ideas.

Applied to the random code ensemble considered here,
the line of thought is as follows: If $\epsilon$ is such that 
$$\Gamma\left(\frac{P_a-\epsilon}{P_g}\right) > - R,$$
then the energy level $\epsilon$ will be typically populated with an exponential
number of codewords, concentrated very strongly around its mean
$$\overline{N(\epsilon)}\exe\exp\left\{n\left[R+
\Gamma\left(\frac{P_a-\epsilon}{P_g}\right)\right]\right\},$$
otherwise (which means that $\overline{N(\epsilon)}$ is exponentially small),
the energy level $\epsilon$ will not be populated by any codewords typically. This
means that the populated energy levels range between
$$\epsilon_1\eqd P_a-P_g\sqrt{1-e^{-2R}}$$
and
$$\epsilon_2\eqd P_a+P_g\sqrt{1-e^{-2R}},$$
or equivalently, the populated values of $\rho$ range between $-\rho_*$ and $+\rho_*$
where $\rho_*=\sqrt{1-e^{-2R}}$. By large deviations and saddle--point methods
\cite{Ellis85, DemZei98},
it follows that for a typical realization of the randomly chosen code, we have
\begin{align*}
Z_e(\beta|\by)
&\exe e^{-nR}\max_{\epsilon\in[\epsilon_1,\epsilon_2]}
\exp\left\{n\left[R+
\Gamma\left(\frac{P_a-\epsilon}{P_g}\right)-\beta\epsilon
\right]\right\}\nonumber\\
&= \max_{\epsilon\in[\epsilon_1,\epsilon_2]}
\exp\left\{n\left[\Gamma\left(\frac{P_a-\epsilon}{P_g}\right)-\beta\epsilon
\right]\right\}\nonumber\\
&= \exp\left\{n\left[\max_{|\rho|\le\rho_*}\left\{\frac12\ln(1-\rho^2)-\beta(P_a-\rho P_g)
\right\}\right]\right\}\ .
\end{align*}
The derivative of $\frac{1}{2}\ln(1-\rho^2)+\rho\beta P_g$ w.r.t.\ $\rho$
vanishes within $[-1,1]$ at:
$$\rho=\rho_\beta\eqd\sqrt{1+\theta^2}-\theta$$
where
$$\theta\eqd\frac{1}{2\beta P_g}.$$
This is the maximizer as long as $\sqrt{1+\theta^2}-\theta\le\rho_*$, namely,
$\theta >e^{-2R}/{2\rho_*}$, or equivalently, $\beta<\rho_* e^{2R}/P_g$, which for
$P_g=\sqrt{P_x(P_x+1/\beta)}$, is equivalent to $\beta<\beta_R\eqd
(e^{2R}-1)/P_x$.
Thus, for the typical code we have
$$\phi_e(\beta,R)\eqd\lim_{n\to\infty}\frac{\ln Z_e(\beta|\by)}{n}=
\begin{cases}
\frac12\ln(1-\rho_\beta^2)-\beta (P_a -\rho_\beta P_g), & \beta < \beta_R\ \\
-R-\beta(P_a-\rho_* P_g), & \beta \ge \beta_R\ .
\end{cases}$$
Taking now into account $Z_c(\beta|\by)$, it is easy to see that for $\beta\ge\beta_R$
(which means $R<C$), $Z_c(\beta|\by)$ dominates $Z_e(\beta|\by)$, whereas for
$\beta<\beta_R$ it is the other way around. It follows then that
$$\phi(\beta,R)\eqd\lim_{n\to\infty}\frac{\ln Z(\beta|\by)}{n}=
\begin{cases}
\frac12 \ln(1-\rho_\beta^2)-\beta (P_a -\rho_\beta P_g), & \beta < \beta_R\\
-R-\frac{1}{2}, & \beta \ge \beta_R\ .
\end{cases}$$
On substituting $P_a=P_x+1/(2\beta)$, $P_g=\sqrt{P_x(P_x+1/\beta)}$ and
$$\rho_\beta=\sqrt{1+\theta^2}-\theta=\sqrt{\frac{\beta P_x}{1+\beta P_x}},$$
we then get:
$$\psi(\beta)=-\lim_{n\to\infty}\frac{\ln Z(\beta|\by)}{n}=
\begin{cases}
\frac{1}{2}\ln(1+\beta P_x)+\frac{1}{2}, & \beta < \beta_R\\
R+\frac{1}{2} & \beta \ge \beta_R\ .
\end{cases}$$
Note that $\psi(\beta)$ is a continuous function but it is not smooth at
$\beta=\beta_R$.
Now,
\begin{equation}  \label{eq:sphere}
\lim_{n\to\infty}\frac{\mbox{mmse}(\bX|\bY)}{n}=2\frac{\mbox{d}\psi(\beta)}{\mbox{d}\beta}
=
\begin{cases}
\frac{P_x}{1+\beta P_x}, & \beta < \beta_R\\
0, & \beta \ge \beta_R  \ .
\end{cases}  
\end{equation}
which means that there is a first order phase transition\footnote{By ``first--order phase
transition'', we mean, in this context, that the MMSE is a discontinuous function of $\beta$.}
in the MMSE: As long as $\beta
\ge \beta_R$, which means $R < C$, the MMSE essentially vanishes since the
correct codeword can be reliably decoded, whereas for $R > C$, the MMSE behaves
as if the inputs were i.i.d.\ Gaussian with variance $P_x$ (cf.\
Subsection \ref{Gaussianiid}).

\subsection{Hierarchical Code Ensemble for the Degraded Broadcast Channel} 
\label{bc}

Consider the following hierarchical code ensemble:
First, randomly draw $M_1=e^{nR_1}$ cloud--center vectors $\{\bu_i\}$ on the 
$\sqrt{n}$--sphere.
Then, for each $\bu_i$, randomly draw $M_2=e^{nR_2}$ codewords
$\{\bx_{i,j}\}$ according to $\bx_{i,j}=\alpha\bu_i+\sqrt{1-\alpha^2}\,\bv_{i,j}$, where
$\{\bv_{i,j}\}$ are randomly drawn uniformly and independently on the
$\sqrt{n}$--sphere.
This means that $\|\bx_{i,j}-\alpha\bu_i\|^2=n(1-\alpha^2)
\eqd nb$.
Without essential loss of generality, 
here and in Subsection \ref{tree}, we take the
channel input power to be $P_x=1$.

Let $\bx_{0,0}$, belonging to cloud center
$\bu_0$, be the
input to the
Gaussian channel~\eqref{GCv}.  It is easy to see that if the SNR of
the Gaussian channel is high enough, the codeword $\bx_{i,j}$ can be
decoded; while at certain lower SNR only the cloud center $\bu_i$ can
be decoded but not $\bv_{i,j}$.  In the following we show the phase
transitions of the MMSE as a function of the SNR.

We will decompose the
partition function as follows:
\begin{align}
Z(\beta|\by)&=e^{-nR}\sum_{i,j}\exp(-\beta\|\by-\bx_{i,j}\|^2/2)\nonumber\\
&=e^{-nR}\exp(-\beta\|\by-\bx_{0,0}\|^2/2)+
e^{-nR}\sum_{j\ge 1}\exp(-\beta\|\by-\bx_{0,j}\|^2/2)\nonumber\\
& \qquad\qquad + e^{-nR}\sum_{i\ge 1}\sum_j\exp(-\beta\|\by-\bx_{i,j}\|^2/2)
\nonumber\\
&\eqd Z_c(\beta|\by)+Z_{e1}(\beta|\by)+Z_{e2}(\beta|\by)
\end{align}
where once again, $Z_c(\beta|\by)$ -- the contribution of the
correct codeword, is typically about $e^{-n(R+1/2)}$. The other two terms
$Z_{e1}(\beta|\by)$ and $Z_{e2}(\beta|\by)$ correspond to contributions of
incorrect codewords from the same cloud and from other clouds, respectively.

Let us consider $Z_{e1}(\beta|\by)$ first. The distance $\|\by-\bx_{0,j}\|^2$
is decomposed as follows:
\begin{align}
\|\by-\bx_{0,j}\|^2
&=\|(\by-\alpha\bu_0)+(\alpha\bu_0-\bx_{0,j})\|^2\nonumber\\
&=\|\by-\alpha\bu_0\|^2+\|\alpha\bu_0-\bx_{0,j}\|^2+2
\langle \by-\alpha\bu_0,\alpha\bu_0-\bx_{0,j} \rangle\ .
\end{align}
Now, $\|\by-\alpha\bu_0\|^2$ is typically about $n/\beta+nb\eqd na$ 
and $\|\alpha\bu_0-\bx_{0,j}\|^2=nb$.
Thus, for $\|\by-\bx_{0,j}\|^2/2$ to be around $n\epsilon$,
$\langle \by-\alpha\bu_0,\alpha\bu_0-\bx_{0,j} \rangle$ 
must be around $n[\epsilon-(a+b)/2]\eqd n[\epsilon-P_a]$.
Now, the question is this: Given $\by-\alpha\bu_0$,
what is the typical number
of codewords in cloud 0 for which
$\langle \by-\alpha\bu_0,\alpha\bu_0-\bx_{0,j} \rangle =n[\epsilon-P_a]$.
Similarly as before, the
answer is the following:
  \begin{equation}    \label{eq:Ne}
N(\epsilon)\exe
\begin{cases}
\exp\left\{n\left[R_2+
\Gamma\left(\frac{\epsilon-P_a}{P_g}\right)
\right]\right\}, &
\epsilon\in[P_a-\rho_2P_g,P_a+\rho_2P_g]\\
0, & \mbox{elsewhere}
\end{cases}
  \end{equation}
where $P_g\eqd\sqrt{ab}$ and $\rho_2=\sqrt{1-e^{-2R_2}}$.
Thus,
\begin{align}
Z_{e1}(\beta|\by)
&\exe e^{-nR}\exp\left\{n\left[\max_{|\rho|\le\rho_2}\left\{
R_2+\Gamma(\rho)-\beta(P_a-\rho P_g)\right\}\right]\right\}\nonumber\\
&= e^{-nR_1}\exp\left\{n\left[\max_{|\rho|\le\rho_2}\left\{
\frac{1}{2}\ln(1-\rho^2)+\beta\rho P_g\right\}-\beta P_a\right]\right\}.
\end{align}
As before, the derivative of $[\frac{1}{2}\ln(1-\rho^2)+\rho\beta P_g]$ w.r.t.\ $\rho$
vanishes within $[-1,1]$ at:
$$\rho=\rho_\beta\eqd\sqrt{1+\theta^2}-\theta$$
where
$$\theta\eqd\frac{1}{2\beta P_g}.$$
This is the maximizer as long as $\sqrt{1+\theta^2}-\theta\le\rho_2$, namely,
$\theta >e^{-2R_2}/{2\rho_2}$, or equivalently, $\beta<\rho_2 e^{2R_2}/P_g$, which for
$P_g=\sqrt{b(b+1/\beta)}$, is equivalent to $\beta<\beta(R_2)\eqd (e^{2R_2}-1)/b$.
Thus, for the typical code we have
$$\psi_{e1}(\beta)\eqd-\lim_{n\to\infty}\frac{\ln Z_{e1}(\beta|\by)}{n}=
\begin{cases}
R_1-\frac{1}{2}\ln(1-\rho_\beta^2)+\beta (P_a -\rho_\beta P_g),
 & \beta < \beta(R_2) \\
R+\beta(P_a-\rho_2 P_g), & \beta \ge \beta(R_2) \ .
\end{cases}$$
Similarly as before, it is easy to see that
$$Z_c+Z_{e1}\exe\exp\left\{-n\left[R_1+\min\left\{R_2,\frac{1}{2}\ln(1+b\beta)\right\}+\frac{1}{2}\right]\right\}.$$
Turning now to $Z_{e2}(\beta|\by)$, we have the following consideration.
Given $\bu_i$, $i\ge 1$, let $\by'=\by-\alpha\bu_i$ and $\bv_{i,j}=\bx_{i,j}-\alpha\bu_i$.
We would like to estimate how many codewords in cloud $i$,
$N_i(\epsilon)$, contribute
$\|\by-\bx_{i,j}\|^2/2=\|\by'-\bv_{i,j}\|^2/2=n\epsilon$.
Similarly as before, $N_i(\epsilon)$ is given by
 exactly the same formula as \eqref{eq:Ne}
where this time,
$P_a=(1-\alpha^2+\|\by-\alpha\bu_i\|^2/n)/2$ and $P_g=\sqrt{(1-\alpha^2)\|\by-\alpha\bu_i\|^2/n}$.
Thus, we have expressed the typical number 
of codewords that cloud $i$ contributes
with energy $\epsilon$ as
$N_i(\epsilon)=\exp\{nF(\|\by-\alpha\bu_i\|^2/n,\epsilon)\}$,
 and the total number is
$N(\epsilon)=\sum_i N_i(\epsilon)$. 
Now let $M(\delta)$ be the number of $\{\bu_i\}$ for
which $\|\by-\alpha\bu_i\|^2/n=\delta$. Then, 
$$N(\epsilon)\exe\sum_\delta M(\delta)e^{nF(\delta,\epsilon)}.$$
Now,
$$M(\delta)=
\begin{cases}
\exp\left\{n\left[R_1+
\Gamma\left(\frac{\delta/2-P_a'}{P_g'}\right)\right]\right\}, &
\delta\in[\delta_1,\delta_2]\ ,\\
0, & \mbox{elsewhere}  \,
\end{cases}$$
where $P_a'=(1+1/\beta+\alpha^2)/2$, $P_g'=\alpha\sqrt{1+1/\beta}$,
$\delta_1=2(P_a'-P_g'\sqrt{1-e^{-2R_1}})\eqd 2(P_a'-\rho_1 P_g')$ and 
$\delta_2=2(P_a'+P_g'\rho_1)$.
Thus,
$$N(\epsilon)\exe\exp\left\{n\max_{\delta_1\le\delta\le\delta_2}
\left[R_1+\Gamma\left(\frac{P_a'-\delta}{P_g'}\right)+
F(\delta,\epsilon)\right]\right\}.$$

\begin{figure}
\begin{center}
\includegraphics[width=4.5in]{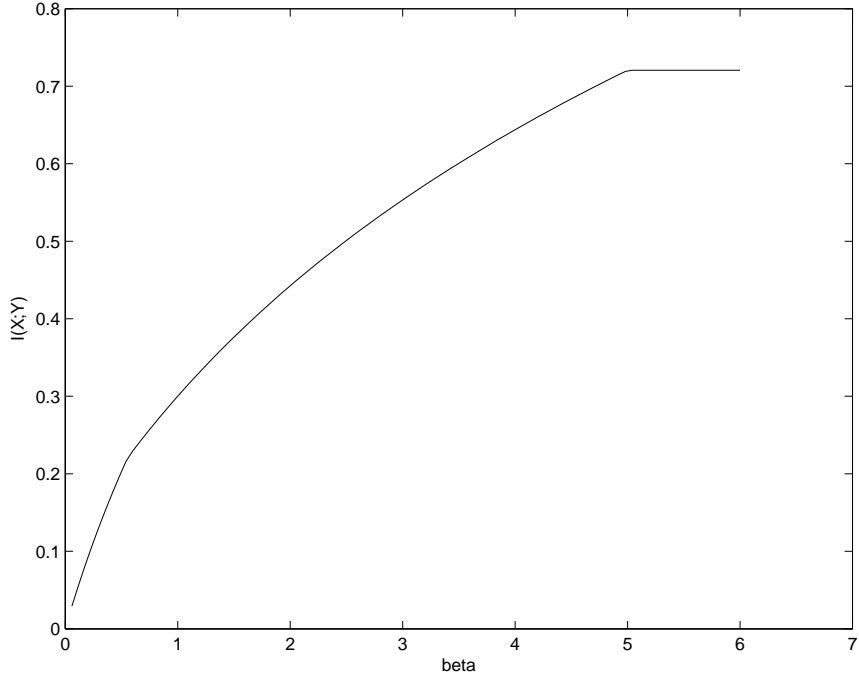}
\caption{Graph of $\lim_{n\to\infty} 
I(\bX;\bY)/n=-\bE_\beta\{\ln Z(\beta|\bY)\}/n-1/2$
as a function of $\beta$
for $R_1=0.1$, $R_2=0.6206$, and $\alpha=0.7129$,
which result in $\beta_1=0.5545$ and $\beta_2=5.001$.
As can be seen quite clearly, there are phase transitions at these values of
$\beta$.}
\label{transforms}
\end{center}
\end{figure}

Putting it all together, we get:
\begin{equation}
  \begin{split}
\psi_{e2}(\beta)
\eqd -\lim_{n\to\infty} & \frac{\ln Z_{e2}(\beta|\by)}{n}
=-\max_{|r_1|\le\rho_1}\;
\max_{|r_2|\le\rho_2(r_1)}\left\{\frac{1}{2}\ln(1-r_1^2)+\frac{1}{2}\ln(1-r_2^2)-\right.\\
& \left.\beta\left[\frac{1-\alpha^2}{2}+P_a'-r_1P_g'-
r_2\sqrt{2(1-\alpha^2)(P_a'-r_1P_g')}\right]\right\},    
  \end{split}
\end{equation}
where $\rho_1=\sqrt{1-e^{-2R_1}}$, 
$\rho_2(r_1)=\sqrt{1-e^{-2R}/(1-r_1^2)}$,
$P_a'=(1+1/\beta+\alpha^2)/2$, and $P_g'=\alpha\sqrt{1+1/\beta}$.
The above expression does not seem to lend itself to closed form analysis in an easy manner.
Numerical results (cf.\ Fig.~1) show a 
reasonable match (within the order of magnitude of $1\times 10^{-5}$)
between values of $\lim_{n\to\infty} I(\bX;\bY)/n$ obtained numerically from the asymptotic exponent
of $\bE_\beta\ln Z(\beta|\bY)$ and those that are obtained from the expected behavior in this case:
$$\lim_{n\to\infty} \frac{I(\bX;\bY)}{n}=
\begin{cases}
\frac{1}{2}\ln(1+\beta), & \beta < \beta_1\\
R_1+\frac{1}{2}\ln(1+\beta b), & \beta_1 \le \beta < \beta_2\\
R=R_1+R_2, & \beta \ge \beta_2  
\end{cases}$$
where
$$\beta_1\eqd \frac{e^{2R_1}-1}{1-be^{2R_1}}\ , 
\quad
\beta_2\eqd \frac{e^{2R_2}-1}{1-b}\ ,$$
and it is assumed that the parameters of the model ($R_1$, $R_2$ and $\alpha$) 
are chosen such that $\beta_1 < \beta_2$. Accordingly, the MMSE undergoes two
phase transitions, where it behaves as if the input was: (i) Gaussian i.i.d.\ with
unit variance for
$\beta < \beta_1$ (where no information can be decoded), (ii) Gaussian input
of a smaller variance (corresponding to the cloud), in the intermediate range (where the cloud center is
decodable, but the refined message is not), and (iii) the MMSE altogether vanishes for $\beta >
\beta_2$, where both messages are reliably decodable.
%

The hierarchical code ensemble takes the superposition code structure
which achieves the capacity region of the Gaussian broadcast channel.
Consider two receivers, referred to as receiver 1 and receiver 2, with
$\beta_1$ and $\beta_2$ respectively.  Receiver 1 can decode the cloud
center, whereas receiver 2 can decode the entire codeword.  In other
words, suppose the hierarchical code ensemble with rate pair $(R_1,
R_2)$ and parameter $\alpha$ is sent to two receivers with fixed SNR
of $\gamma_1$ and $\gamma_2$ respectively.  Then the minimum decoding
error probability vanishes as long as $(R_1,R_2,\alpha)$ are such that
\begin{align}
  R_1&<\frac12\log\left(1+\frac{\alpha^2 \gamma_1}{1+(1-\alpha^2)\gamma_1} \right), 
  \label{eq:bc1} \\
  R_2&<\frac12\log\left( 1 + \alpha^2 \gamma_2 \right). \label{eq:bc2}
\end{align}
In particular, all boundary points of the capacity region can be
achieved by varying the power distribution coefficient $\alpha$.
This capacity region result also leads
to the fact that if only the cloud center is decodable, then the MMSE
for the codeword $\bv_{i,j}$ is no different to that if the elements
of $\bv_{i,j}$ were i.i.d.\ standard Gaussian.  Knowledge of the
codebook structure of $\{\bv_{i,j}\}$ does not reduce the MMSE because
otherwise the code cannot achieve the capacity region of the Gaussian
broadcast channel.

\subsection{Hierarchical Tree--Structured Code} 
\label{tree}

Consider next an hierarchical code with the following 
structure: The block of length $n$ is partitioned into two segments,
the first is of length $n_1=\lambda_1n$ ($\lambda_1\in(0,1)$) and the
second is of length $n_2=\lambda_2n$ ($\lambda_2=1-\lambda_1$). We
randomly draw $M_1=e^{n_1R_1}$ first--segment
codewords $\{\bx_i\}$ on the surface of
the $\sqrt{n_1}$--sphere, and then, for each $\bx_i$, we randomly
draw $M_2=e^{n_2R_2}$ second--segment codewords $\{\bx_{i,j}'\}$
on the surface of
the $\sqrt{n_2}$--sphere. The total message of length $nR=n_1R_1+n_2R_2$
(thus $R=\lambda_1R_1+\lambda_2R_2$)
is encoded in two parts: The first--segment codeword depends only on the
first $n_1R_1$ bits of the message whereas the second--segment codeword
depends on the entire message. 

Let $(\bx_0,\bx_{0,0})$ be the transmitted codeword, and
let $\by$ and $\by'$ be the corresponding segments of the channel
output vector $(\by,\by')$.
The partition function is as follows:
\begin{align}
Z(\beta|\by)&=e^{-nR}\exp\{-\beta[\|\by-\bx_0\|^2+\|\by'-\bx_{0,0}\|^2]/2\}
\nonumber\\
&\qquad +e^{-nR}\exp\{-\beta[\|\by-\bx_0\|^2/2\}\sum_j
\exp\{-\beta\|\by'-\bx_{0,j}\|^2]/2\}\nonumber\\
& \qquad +e^{-nR}\sum_{i\ge 1}\sum_j\exp\{-\beta[\|\by-\bx_i\|^2/2\}
\exp\{-\beta\|\by'-\bx_{i,j}\|^2]/2\}\nonumber\\
&\eqd Z_c+Z_{e1}+Z_{e2}.
\end{align}
Now, as before, $Z_c\exe e^{-n(R+1/2)}$.
As for $Z_{e1}$, it can also be treated as in Subsection \ref{simplecode}:
The first factor contributes $e^{-nR}\cdot e^{-n\lambda_1/2}$.
The second factor is $e^{-n\lambda_2[\min\{R_2,C(\beta)\}+1/2]}$,
where $C(\beta)=\frac{1}{2}\ln(1+\beta)$.
Thus,
$$Z_{e1}(\beta|\by)+Z_c\exe
\exp\left\{-n\left[\lambda_1R_1+\
\lambda_2\min\{R_2,C(\beta)\}+\frac{1}{2}\right]\right\}.$$
Consider next the term $Z_{e2}$. Let 
$r_1=\langle\bx,\by\rangle/(n_1P_g)$ 
and
$r_2=\langle\bx',\by'\rangle/(n_2P_g)$
where $P_g$ is as in Subsection \ref{simplecode}.
Of course,
$\langle(\bx,\bx'),(\by,\by')\rangle /(nP_g)=\lambda_1r_1+\lambda_2r_2$.
What is the typical number of codewords 
$(\bx_i,\bx_{i,j}')$ of $Z_{e2}$ whose correlation with $(\by,\by')$
is exactly $r$? The answer is
$$\lim_{n\to\infty}\frac{\ln N(r)}{n}=\max_{|r_1|\le\rho(R_1)}
\left\{\lambda_1R_1+{\lambda_1}\Gamma(r_1) + 
\lambda_2R_2+{\lambda_2}\Gamma
\left(\frac{r-\lambda_1r_1}{\lambda_2}\right)\right\},$$
where $\rho(x)=\sqrt{1-e^{-2x}}$.
This expression behaves differently depending on whether
$R_1 > R_2$ or $R_1 < R_2$. In the first case, it behaves exactly as
in the ordinary ensemble, that is:
$$\lim_{n\to\infty}\frac{\ln N(r)}{n}=
\begin{cases}
R+\frac{1}{2}\ln(1-r^2), & |r|\le\rho(R)\\
0, & |r|>\rho(R)  \ .
\end{cases}
$$
and then, of course, $Z_{e2}$ is as before:
$$Z_{e2}+Z_c\exe \exp\{-n[\min\{R,C(\beta)\}+1/2]\}.$$
When $R_1 < R_2$, however, we have two phase transitions:
$$\lim_{n\to\infty}\frac{\ln N(r)}{n}=
\begin{cases}
R+\Gamma(r), & |r|\le\rho(R_1)\\
\lambda_2\left[R_2+\Gamma\left(\frac{r-\lambda_1\rho(R_1)}{\lambda_2}\right)\right],
& \rho(R_1)\le |r|\le\lambda_1\rho(R_1)+\lambda_2\rho(R_2)\\
0, & |r|>\lambda_1\rho(R_1)+\lambda_2\rho(R_2)\ .
\end{cases}
$$
In this case, we get:
$$\lim_{n\to\infty}\frac{\ln(Z_{e2}+Z_c)}{n}=
\begin{cases}
-C(\beta)-\frac{1}{2}, & \beta \le \beta(R_1)\\
-\lambda_1R_1-\lambda_2C(\beta)-\frac{1}{2}, & \beta(R_1)<\beta\le\beta(R_2)\\
-R-\frac{1}{2}, & \beta > \beta(R_2)  
\end{cases}
$$
where $\beta(R)$ is the solution $\beta$ to the equation $C(\beta)\equiv
\frac{1}{2}\ln(1+\beta)=R$.
To summarize, we have the following: $Z_c\exe e^{-n(R+1/2)}$,
$Z_{e1}+Z_c\exe \exp\{-n[\lambda_1R_1+\lambda_2\min\{R_2,C(\beta)\}+1/2]\}$
and 
$$Z_{e2}+Z_c\exe
\begin{cases}
\exp\{-n[\min\{R,C(\beta)\}+1/2]\}, & R_1>R_2\\
\exp\{-n[\lambda_1\min\{R_1,C(\beta)\}+\lambda_2\min\{R_2,C(\beta)\}+1/2]\} ,
& R_1\le R_2  \ .
\end{cases}
$$
Clearly, if $R_1\le R_2$ then $Z_{e2}+Z_c$ dominates $Z_{e1}+Z_c$.
If $R_1>R_2$, we note that
$$\min\{\lambda_1 R_1+\lambda_2 \min\{R_2,C(\beta)\}, \min\{R,C(\beta)\}\}\equiv
\min\{R,C(\beta)\}.$$
Thus,
$$Z\exe
\begin{cases}
\exp\{-n[\min\{R,C(\beta)\}+1/2]\} ,
& R_1>R_2\\
\exp\{-n[\lambda_1\min\{R_1,C(\beta)\}+\lambda_2\min\{R_2,C(\beta)\}+1/2]\} ,
& R_1\le R_2  \ .
\end{cases}
$$
The MMSE then is as in \eqref{eq:sphere} in Subsection \ref{simplecode} when $R_1>R_2$, and
given by
\begin{equation}  \label{eq:tree}
\mbox{mmse}(\bX|\bY)=
\begin{cases}
\frac{1}{1+\beta}, & \beta \le\beta(R_1)\\
\frac{\lambda_2}{1+\beta}, & \beta(R_1)< \beta \le\beta(R_2)\\
0, & \beta > \beta(R_2)  
\end{cases}
\end{equation}
when $R_1<R_2$. This dichotomy between these two types of behavior
have their roots in the behavior of the GREM, a generalized version of the random energy
model, where the random energy levels of the various system
configurations are correlated (rather than being i.i.d.) in an hierarchical
structure
\cite{Derrida85, DG86a, DG86b}. The GREM turns out to
have an intimate analogy with the tree--structured code ensemble considered here.
The reader is referred to \cite{Merhav09} for a more elaborate discussion
on this topic.

The preceding result on the MMSE is consistent with the analysis based
solely on information theoretic considerations.  In case $R_1<R_2$,
the first segment code is decodable as long as $R_1
<(1/2)\log(1+\beta)$, whereas the second segment code is decodable if
also $R_2 < (1/2)\log(1+\beta)$.  Hence the MMSE is given by
\eqref{eq:tree}.
In case $R_1>R_2$, the second-segment code is decodable if and only if
the first-segment is also decodable, i.e., the two codes can be
decoded jointly.  This requires $R_2 < (1/2) \log(1+\beta)$,
$\lambda_1 R_1 < \lambda_1 \log(1+\beta) + \lambda_2 \log(1+\beta)$
and $R = \lambda_1 R_1 + \lambda_2 R_2 < \log(1+\beta)$.  The last
inequality dominates, hence the MMSE is given by \eqref{eq:sphere}.

\subsection{Estimation of Sparse Signals}
\label{sparse}

Let the components
of $\bX$ be given by $X_i=S_iU_i$, 
$i=1,2,\ldots,n$, where $S_i\in\{0,1\}$
and $\{U_i\}$ are $\calN(0,\sigma^2)$ i.i.d.\ and independent of $\{X_i\}$. 
As before $\bY=\bX+\bN$, where
the components of $\bN$ are i.i.d.\ Gaussian $\calN(0,1/\beta)$.
One motivation of this simple model is in compressed sensing applications, where 
the signal $\bX$ (possibly, in some transform domain) is assumed to possess
a limited fraction of non--zero components,
here designated by the non--zero components of $\bS=(S_1,S_2,\ldots,S_n)$. The signal $\bX$ is
considered sparse if the relative fraction of 1's in $\bS$ is small. We will assume that $\bS$, whose
realization is not revealed to the estimator, is
governed by a given probability distribution $P(\bs)$. We first derive an expression of the
partition function for a general $P(\bs)$ and then particularize our study to a certain form
of $P(\bs)$. First, we have the following:
\begin{align}
P(\bx)&=\sum_{\bs}P(\bs)P(\bx|\bs)\nonumber\\
&=\sum_{\bs}P(\bs)
\prod_{i:~s_i=0}\delta(x_i)\prod_{i:~s_i=1}
\left[(2\pi\sigma^2)^{-1/2}\exp\{-x_i^2/(2\sigma^2)\}\right]\nonumber\\
&=\sum_{\bs}P(\bs)
\prod_{i=1}^n
\left[(2\pi s_i\sigma^2)^{-1/2}\exp\{-x_i^2/(2s_i\sigma^2)\}\right]
\label{eq:Px}
\end{align}
where a zero--variance Gaussian distribution is understood to be
equivalent to the Dirac delta--function. Thus,
\begin{align}
Z(\beta|\by)&=\int_{\reals^n}d\bx P(\bx)\exp\{-\beta\|\by-\bx\|^2/2\}\nonumber\\
&=\sum_{\bs}P(\bs)\prod_{i=1}^n\left[\int_{-\infty}^\infty \intd x_i
(2\pi s_i\sigma^2)^{-1/2}\exp\{-x_i^2/(2s_i\sigma^2)\}\cdot\exp\{-\beta(y_i-x_i)^2/2\}\right]
\nonumber\\
&=\sum_{\bs}P(\bs)\prod_{i=1}^n\left[(1+qs_i)^{-1/2}
\exp\left\{-\frac{\beta y_i^2}{2(1+qs_i)}\right\}\right]\nonumber\\
&=\sum_{\bs}P(\bs)\prod_{i=1}^n
\exp\left\{-\frac{1}{2}\left[\frac{\beta y_i^2}{1+qs_i}+
\ln(1+qs_i)\right]\right\}
\end{align}
where we have used the
notation\footnote{The quantity $q$ is proportional to the SNR.}
$q=\beta\sigma^2$.
Transforming $\bs$ to ``spins'' 
$\bmu=(\mu_1,\ldots,\mu_n)$ by the relation $\mu_i=1-2s_i\in\{-1,+1\}$, we get:
$$\frac{\beta y_i^2}{1+qs_i}+\ln(1+qs_i) =
\frac{(1+q/2)\beta y_i^2}{1+q}+\frac{1}{2}\ln(1+q)
-2 \mu_i h_i$$
where
\begin{equation}  \label{eq:hi}
  h_i=-\frac{\beta^2\sigma^2 y_i^2}{4\big(1+\beta\sigma^2\big)}
  +\frac{1}{4}\ln\big(1+\beta\sigma^2\big).
\end{equation}
On substituting back into the partition function we get:
\begin{equation}
\label{genZ}
Z(\beta|\by)=(1+q)^{-n/4}\cdot
\exp\left\{-\frac{\beta(1+q/2)}{2(1+q)}\|\by\|^2\right\}\cdot
\sum_{\bmu} P(\bmu)\exp\left\{\sum_{i=1}^n\mu_ih_i\right\}.
\end{equation}
Thus $h_i$
is given the statistical--mechanical
interpretation of the random `local' magnetic field felt by the $i$--th spin.

Eq.\ (\ref{genZ}) holds for a general distribution 
$P(\bs)$ or equivalently, $P(\bmu)$. To further develop
this expression, we must make some assumptions on one of these distributions. 
At this point, we have the freedom to examine 
certain models of $P(\bmu)$, and by viewing the expression
$\sum_{\bmu} P(\bmu)\exp\{\sum_i\mu_ih_i\}$ 
as the partition function
of a certain spin system with a non--uniform, random field $\{H_i\}$ (whose
realization is $\{h_i\}$),
we can borrow techniques from statistical 
physics to analyze its behavior. Evidently,
for every spin glass model that exhibits phase 
transitions, it is conceivable that there will
be analogous phase transitions in the 
corresponding signal estimation problem. 

Assuming certain symmetry properties among the 
various components of $\bs$, it would be plausible to postulate
that all $\{\bs\}$ with the same
number of 1's are equally likely, or equivalently,
all spin configurations $\{\bmu\}$ with the same magnetization
$$m(\bmu)=\frac{1}{n}\sum_{i=1}^n \mu_i$$
have the same probability. 
This means that $P(\bmu)$ depends on $\bmu$ only via $m(\bmu)$.
Consider then the form
$$P(\bmu)=C_n\exp\{nf(m(\bmu))\},$$
where $f(m)$ is an arbitrary function and $C_n$ is a normalization constant.
Further, let us assume that $f$ is twice differentiable
with finite first derivative on $[-1,1]$.
Clearly,
\begin{align}
  C_n&=\bigg( \sum_{\bmu}\exp\{n\,f(m(\bmu))\} \bigg)^{-1} \nonumber\\
  &\exe \exp\left\{ -n\max_m\{ \Hb((1+m)/2)+f(m) \} \right\} \nonumber\\
  &=\exp\left\{ -n\left( \Hb((1+m_a)/2)+f(m_a) \right) \right\}
  \label{eq:Cn}
\end{align}
where $\Hb(\cdot)$ denotes the binary entropy function
and 
$m_a$ is the maximizer of $\Hb((1+m)/2)+f(m)$.
In other words,
$m_a$ is the {\it a--priori} magnetization, namely the magnetization
that dominates $P(\bmu)$.
Of course, when $f(m)$ is linear in $m$, 
the components of $\bmu$ are i.i.d.
Note that if $f$ is monotonically increasing in $m$,
then $P(\bmu)$ has a sharp peak at $m=1$, which
corresponds to a vanishing fraction of 
sites with $s_i=1$, i.e., a sparse signal.
Our derivation, however, will take place for general $f$.

\subsubsection{General Solution}

On substituting the above expression of $P(\bmu)$ into that of $Z(\beta|\by)$,
our main concern is then how to deal with the expression
\begin{equation}  \label{eq:Zbh}
\hat{Z}(\beta|\bh)\eqd\sum_{\bmu}P(\bmu)e^{\sum_i\mu_ih_i}=
C_n\sum_{\bmu}\exp\left\{n\left[f(m(\bmu))+
\frac{1}{n}\sum_i\mu_ih_i\right]\right\}.  
\end{equation}
We investigate the typical behavior of the partition function, or more
precisely, calculate the following quantity:
\begin{equation}  \label{eq:Zh}
\frac1n \log \bE\left\{ \hat{Z}(\beta|\bH) \right\}
= \frac1n \log \left[
C_n \bE\left\{ \sum_{\bmu}\exp\left\{n\left[f(m(\bmu))+
\frac{1}{n}\sum_i\mu_iH_i\right]\right\}\right\}\right]  
\end{equation}
where $\bH$ consists of i.i.d.\ random variables with arbitrary
distribution $p(H)$.

Using large deviations theory, as $n\to\infty$, the dominant value of
$m$ in \eqref{eq:Zh}, henceforth denoted as $m^*$ is shown to satisfy
\begin{equation}
\label{mstarequation}
m^*=\bE\{\tanh(f'(m^*)+H)\}
\end{equation}
and
\begin{equation}  \label{eq:AT}
\bE\{\tanh^2(f'(m^*)+H)\}> 1-\frac{1}{f''(m^*)}.
\end{equation}
The detailed analysis is relegated to Appendix \ref{app:m}.
Clearly, $m^*$ is the dominant magnetization {\it a--posteriori},
i.e., the one that dominates the posterior of $m(\bmu)$ given (a
typical) $\by$.
It is also shown in Appendix \ref{app:m} that
\begin{equation}
  \lim_{n\to\infty} \frac1n \log \bE\left\{ \hat{Z}(\beta|\bH) \right\}
  = \lim_{n\to\infty} \frac1n \log C_n - \psi(m^*)
\end{equation}
where 
\begin{equation} \label{eq:psistar}
  \psi(m^*) \eqd f'(m^*)\, m^* -  f(m^*) 
  - \bE\left\{ \log \left[2 \cosh( f'(m^*) + H)\right] \right\}
\end{equation}
and the normalized exponent of $C_n$ is given by \eqref{eq:Cn}.
Thus the asymptotic normalized mutual information is expressed as
\begin{equation} \label{eq:I}
  \lim_{n\to\infty}\frac{I(\bX;\bY)}{n}=
-\frac{1}{2}+\frac{1}{4}\ln(1+q)+
\frac{\beta(1+q/2)\bE\{Y^2\}}{2(1+q)}-\lim_{n\to\infty}\frac{\ln C_n}{n}
+ \psi(m^*).
\end{equation}
For the sparse signal model described by \eqref{eq:Px},
$H$ is defined by \eqref{eq:hi} with $y_i$ replaced by $Y$
and the expectation over $Y$ is w.r.t.\ a mixture of two Gaussians:
$\calN(0,1/\beta)$ with weight $(1+m_a)/2$, and 
$\calN(0,\sigma^2+1/\beta)$ with weight $(1-m_a)/2$.

The solution to
\begin{equation}  \label{eq:crit}
\bE\{\tanh^2(f'(m)+H)\}= 1-\frac{1}{f''(m)}
\end{equation}
is known as a {\it critical point}, beyond which the solution to
\eqref{mstarequation} ceases to be 
a local maximum and it becomes a local minimum. 
The dominant $m^*$ must jump elsewhere.
Also, as we vary one of the other parameters of the 
model, it might happen that the global maximum
jumps from one local maximum to another.

\subsubsection{Special Case with Quadratic Exponent}

In the case where $f$ is quadratic\footnote{A quadratic model can be thought of 
as consisting of the first few terms of the Taylor expansion of a smooth function $f$.}
in $m$, i.e., 
\begin{equation} \label{quadratic}
f(m)=am+bm^2/2.
\end{equation}
This is similar though not identical
to the {\it random--field Curie-Weiss model} (RFCW model) of spin systems%
\footnote{There is a certain difference in the sense that 
in the RFCW
$\{H_i\}$ 
are i.i.d.,
whereas here each
 $H_i$ 
depends on 
the corresponding $\mu_i$ because the variance of $y_i$ depends on whether
$\mu_i=-1$ or $\mu_i=+1$. Also as a result, 
$\{H_i\}$ 
here are not i.i.d.\ because they depend
on each other via the dependence between $\{\mu_i\}$.
These differences are not crucial, however.}
(cf.\ e.g., \cite{ABI08} and references therein).  
 Eq.\ (\ref{mstarequation}) becomes
$$m=\bE\{\tanh(bm+a+H)\},$$
similarly as in the mean field model with a random field \cite{ABI08}.
Eq.\ \eqref{eq:crit} for the critical point 
satisfies
\begin{equation}  \label{eq:crit2}
  \bE\{\tanh^2(bm+a+H)\} = 1-(1/b).
\end{equation}
 
To demonstrate that 
the global maximum might jump from one local maximum to another,
consider the quadratic case and assume that $\beta$ and $\sigma^2$
are so small that the fluctuations in $H$ can be neglected. Equation
(\ref{mstarequation})
can then be approximated by 
$$m=\tanh(bm+a),$$
which is actually the same the equation of the magnetization as in the 
Curie--Weiss model (a.k.a.\ the mean field model or the infinite--range model)
of spin arrays (cf.\ e.g., \cite[Sect.\ 4.2]{NO88},
\cite[Chap.\ 3]{Baxter82}, \cite[Sect.\ 4.5.1]{Honerkamp02}), which is
actually a special case of the above with 
$H_i\equiv 0$ 
for all $i$.
For $a=0$ and $b> 1$, this equation has two 
symmetric non--zero solutions $\pm m_0$, which both dominate
the partition function. If $a\ne 0$ but small, then the symmetry is broken, and 
there is only one dominant solution which is
about $m_0\,\mbox{sgn}(a)$. To approximate $m_0$ for the case where $|a|$ is small 
and $b$ is only slightly larger than $1$, one can use the
Taylor expansion of the function $\tanh(\cdot)$ (as is customarily done 
in the theory of the infinite range Ising model;
see e.g., \cite[p.\ 188, eqs.\ (4.21a), (4.21b)]{NO88}) and get
$$m\approx bm+a-\frac{(bm+a)^3}{3}.$$
Neglecting the contribution of $a$, we get a simple quadratic equation whose
solutions are $\pm m_0$ with $m_0=\frac{1}{b}\sqrt{3(1-1/b)}$. Thus,
for small values of $|a|$ and $b-1$, 
$$m^*\approx m_0\cdot\mbox{sgn}(a),$$
and so, $m^*$ jumps between $+m_0$ and $-m_0$ as $a$ crosses the origin.
Similarly, for $a=0$, 
$m^*$ jumps from zero to $+m_0$ or $-m_0$ as $b$ passes the value $b=1$ while increasing.

By \eqref{eq:I},
the asymptotic normalized mutual information of this model is given by
\begin{align}
\label{mutualinfo}
\lim_{n\to\infty}\frac{I(\bX;\bY)}{n}
&=-\frac{1}{2}+\frac{1}{4}\ln(1+q)+
\frac{\beta(1+q/2)}{2(1+q)}
\left[\frac{1+m_a}{2}\cdot\frac{1}{\beta}+
\frac{1-m_a}{2}\left(\sigma^2+\frac{1}{\beta}\right)\right]\nonumber\\
& \qquad + \Hb\left(\frac{1+m_a}{2}\right)+f(m_a)
+ \psi(m^*)
\nonumber\\
&=-\frac{1}{2}+\frac{1}{4}\ln(1+q)+
\frac{1+q/2}{2(1+q)}
\left(1+\frac{1-m_a}{2}\cdot q\right)
 + \Hb\left(\frac{1+m_a}{2}\right) \nonumber\\
& \qquad
+ a m_a + \frac{b m_a^2}{2}
-\bE\{\ln[2\cosh(bm^*+a+H)]\}+\frac{b(m^*)^2}{2}.
\end{align}
In this special case of quadratic exponent, the Hubbard-Stratonovich
transformation can be used to obtain an alternative, more
straightforward derivation of the mutual information result
\eqref{mutualinfo}.  The details are provided in Appendix \ref{app:I}.

The MMSE is equal to twice the derivative of (\ref{mutualinfo}) 
w.r.t.\ $\beta$. 
Note that the dominant value $m^*$ is dependent on $\beta$.  In
Appendix \ref{app:E}, we carry out the calculation and obtain
\begin{equation} \label{eq:mmse}
  \begin{split}
    &\lim_{n\to\infty} \frac{\mbox{mmse}(\bX|\bY)}{n} \\
    &\quad= \frac{\sigma^2q}{2(1+q)^2}+\frac{(1-m_a)\sigma^2}{2}
    \left[1-\frac{q(1+q/2)}{(1+q)^2}\right]\\
    &\quad\;\; +
    \frac{1+m_a}{2}\bigg[\mbox{Cov}_0\{Y^2,\ln[2\cosh(bm^*+a+H)]\}
    + \bE_0\{H'\tanh(bm^*+a+H)\}\bigg]\\
    &\quad\;\; + \frac{1-m_a}{2}\bigg[\frac{1}{(1+q)^2}\cdot
    \mbox{Cov}_1\{Y^2,\ln[2\cosh(bm^*+a+H)]\}
    +\bE_1\{H'\tanh(bm^*+a+H)\}\bigg]
  \end{split}
\end{equation}
where $H'$ is defined by
\begin{equation}  \label{eq:hip}
H' = -\frac{\sigma^2}{2(1+q)}+
\frac{q(q+2)}{2(1+q)^2}\cdot Y^2
\end{equation}
which is in fact the derivative of \eqref{eq:hi} w.r.t.~$\beta$.
To ease understanding of the MMSE, we evaluate its value in two
extreme cases in Appendix \ref{app:ex}.

\subsubsection{Discussion}

Returning now to the general expression of the MMSE,
it is reasonable to expect that at the critical points, where $m^*$ jumps from one solution of
eq.\ (\ref{mstarequation}) to another as the
parameters of the model vary, the MMSE may also undergo
an abrupt change, and so the MMSE may be discontinuous (w.r.t.\ these parameters) at these points.
A related 
abrupt change takes place also
in the {\it response} of the MMSE estimator itself at the critical points: Note
that $m^*$ is the dominant magnetization {\it a--posteriori}.
Thus, as $m^*$ jumps,
say, from $m^*=m_1$ to $m^*=m_2$, the
conditional mean estimator, which is a weighted average of $\{\bx\}$,
transfers most of the weight from
a set of $\bx$--vectors
whose binary support vectors $\{\bs\}$ correspond to magnetization $m_1$,
into another set of
$\bx$--vectors supported by $\{\bs\}$ with magnetization $m_2$.
It is not surprising then that this abrupt change in the response of the
estimator is accompanied by a corresponding sudden drop in the MMSE.

It is instructive to compare the type of the phase transition in our example
to those of the ordinary Curie--Weiss model. In the Curie-Weiss model, we have:
\begin{itemize}
\item
A first order phase transition w.r.t.\ the magnetic field
(below the critical temperature), i.e.,
the first derivative of the free energy w.r.t.\
the magnetic field (which is exactly the magnetization) is
discontinuous (at the point of zero field).
\item A second order phase transition w.r.t.\ temperature, i.e.,
the first derivative of the free energy w.r.t.\
temperature (which is related to the internal energy) is
continuous, but the second derivative (which is related to the
specific heat) is not.
\end{itemize}
Here, on the other hand, in physics terms, 
what we observe is a first order phase
transition w.r.t.\ temperature. The reason for this discrepancy is that
in our model, the dependency of the 
free energy on temperature is introduced via the variables $\{h_i\}$ that play
the role of magnetic fields.

In case of quadratic exponent \eqref{quadratic},
$b=0$ corresponds to the special case of i.i.d.\ $\{S_i\}$.  In
this case, our problem is analogous to a system of non-interacting
particles, where of course, no phase transitions can exist. Therefore,
what we learn from statistical physics here is that phase transitions
in the MMSE estimator cannot be a property of the sparsity alone
(because sparsity may be present also for the i.i.d. case with
$P\{S_i=1\}$ small), but rather a property of strong dependency
between $\{S_i\}$, whether it comes with sparsity or not.

\section*{Acknowledgement}
N.~Merhav would like to thank Yonina Eldar for a few interesting discussions
concerning the example of estimating sparse signals (in Subsection
\ref{sparse}) during the early stages
of this work.

\section*{Appendix A -- Estimation of Sparse Signals: The Dominant Magnetization}
\label{app:m}

For the time being let us assume that $H_i$, $i=1,\dots,n$ take on
values from a discrete set $\{h_1,\dots,h_K\}$, where of the $n$
variables, $q_k n$ of them taking the value of $h_k$.  The sum in
\eqref{eq:Zh} can be rewritten as
\begin{equation} \label{eq:syx1}
  \sum_{\bmu} \exp \left\{n f(m(\bmu)) +  \sum^K_{k=1} h_k \sum^{q_kn}_{i=1} \mu_{ki}\right\}
\end{equation}
where we relabel $\mu_i$ as $\mu_{ki}$ with $i=1,\dots,q_kn$ for each $k$.
The expectation on the r.h.s.~of \eqref{eq:Zh} can be viewed as an integral
\begin{equation} \label{eq:inte}
  2^n \int^1_{-1}\cdots \int^1_{-1} \exp\left\{n f(m) + \sum^K_{k=1} h_k (q_kn)
    m_k\right\} N(
  \intd m_1, \cdots, \intd m_K )
\end{equation}
where 
$N$ is a probability measure
proportional to the number of sequences $\bmu$ with $\frac1{q_k n}
\sum^{q_kn}_{i=1} \mu_{ki} \approx m_k$.  
Here $m = \sum^K_{k=1} q_k m_k$.
For $\bmu$ uniformly
randomly chosen from $\pm1$ sequences, the probability measure satisfies
large deviations property, the rate function (or entropy) of which is
obtained as (using the Legendre-Fenchel transform)\footnote{By
  Cram\'er's theorem \cite[Theorem II.4.1]{Ellis85}, the probability
  measure of the empirical mean $\frac1n X_i$ of i.i.d.\ random
  variables $X_i$ satisfy, as $n\rightarrow\infty$, the large
  deviations property with some rate function $I(m)$.  The rate
  of the probability measure is given by the Legendre-Fenchel
  transform\index{Legendre-Fenchel transform} of the cumulant
  generating function (logarithm of the moment generating function)
  \cite{Ellis85, DemZei98}:
  \begin{equation}  \label{e:iuq}
    I(m) = \sup_{\eta} \left[ \eta\,m - \log \bE\left\{e^{\eta X}\right\}
    \right].
  \end{equation}
  It is straightforward to generalize to the product measure of the
  means of subgroups of i.i.d.\ random variables.}
\begin{equation}  \label{eq:Ik}
  I(m_1,\dots,m_K)  = \sum^K_{k=1} q_k
  \left( \log 2 - \Hb\left( \frac{1+m_k}{2}    \right) \right).
\end{equation}
Not surprisingly, the rate function achieves its maximum at
$m_k=0$, $k=1,\dots,K$, where the number of $\pm 1$'s in each
subsequence $\mu_{ki}$, $i=1,\dots,q_kn$ is balanced.
Due to large deviations property, the integral \eqref{eq:inte} is
dominated by unique values of $m_k$, $k=1,\dots,K$.
Specifically, we use Varadhan's Theorem \cite{Ellis85, DemZei98} to
obtain\footnote{The Varadhan's Theorem basically states that, if the
  sequence of probability measures $N_n$ on $\reals$ satisfies large
  deviations property with rate function $I(m)$, and that $F$ is
  continuous and upper bounded on $\reals$, then
  \begin{equation}
    \lim_{n\to\infty} \frac1n \log \int_\reals \exp\{F(m)\} N_n(\intd m)
    = \sup_m \{ F(m) - I(m) \}\ .
  \end{equation}
  The result can also be generalized to multiple dimensions.}
\begin{align}
  \frac1n \log & \int\cdots\int \exp\left\{ n f(m) +  \sum^K_{k=1} h_k (q_kn) m_k \right\}
  N(\intd m_1,\dots,\intd m_k) \nonumber \\
  &\rightarrow \sup_{m_1,\dots,m_K\in[-1,1]}  \left\{ f(m) + \sum^K_{k=1} h_k
  q_k m_k - I(m_1,\dots,m_K) \right\}  \nonumber\\
&= 2^{-n} \cdot \sup_{m_1,\dots,m_K\in[-1,1]}  \psi(m_1,\dots,m_K)
\end{align}
where we use \eqref{eq:Ik} and define
\begin{equation} \label{eq:psik} %
  \psi(m_1,\dots,m_K) \eqd f\left(\sum^K_{k=1} q_k m_k \right) +
  \sum^K_{k=1} h_k q_k m_k + \sum^K_{k=1} q_k \Hb\left(
    \frac{1+m_k}{2} \right).
\end{equation}

The maximum of $\psi$ is achieved by an internal point in $(-1,1)^K$.
This is because $\Hb$ is concave with infinite derivative at the
boundary $m_k=\pm 1$, whereas the derivative of $f$ is finite by
assumption.  Because the function $\psi$ is twice differentiable, at
its maximum, the gradient of $\psi$ w.r.t.~every $m_k$ should be equal
to 0, whereas the Hessian of $\psi$ should be negative definite.
%
It can be shown by taking derivative of $\psi$ w.r.t.\ $m_k$ that zero
gradient is achieved by setting
\begin{equation} \label{eq:mk}
  m_k = \tanh\left( f'\left( \sum^K_{l=1} q_l m_l \right) + h_k \right)
\end{equation}
for all $k$, so that
\begin{equation} \label{eq:m}
  m = \sum^K_{k=1} q_k \tanh\left(f'(m)+h_k\right).
\end{equation}
The Hessian of $\psi$ is determined by noting that
\begin{equation}
  \frac{\partial^2 \psi}{\partial m_k \partial m_l}
  = q_k q_l f''(m)
  - q_k \frac{\delta_{k,l}}{1-m_k^2}
\end{equation}
where $\delta_{k,l}$ is equal to 1 if $k=l$ and equal to 0 otherwise.
The Hessian is negative definite if and only if
\begin{equation}
  \left( \sum^K_{k=1} q_k x_k \right)^2  f''(m)  \le
  \sum^K_{k=1} q_k \frac{x_k^2}{1-m_k^2}
\end{equation}
for all $x_k\in\reals$, $k=1,\dots,K$, which is equivalent to
\begin{equation} \label{eq:f2min}
  f''(m) \le \min_{x_1,\dots,x_K} 
  \frac{ \sum^K_{k=1} q_k x_k^2 / (1-m_k^2)
  }{ \left(\sum^K_{k=1} q_k x_k \right)^2 }.
\end{equation}
Using Lagrange multiplier, the minimum on the r.h.s.\ of
\eqref{eq:f2min} is
obtained as $1-\sum^K_{k=1} q_k m_k^2$.  Further, by~\eqref{eq:mk},
the condition \eqref{eq:f2min} reduces to
\begin{equation} \label{eq:ATk}
  f''(m) \le \frac1{1-\sum^K_{k=1} q_k \tanh^2(f'(m)+h_k)}.
\end{equation}
In other words, a solution of \eqref{eq:mk} is a local
maximum of $\psi$ if and only if it also satisfies \eqref{eq:ATk}.  In
multiple such solutions exist, the global supremum is identified by
comparing the corresponding values of $\psi$.

In the limit $n\to\infty$, the requirement that $H_i$ take
discrete values is not necessary (the continuous distribution can be
regarded as the limit of a degenerate discrete one).  Using
\eqref{eq:m} and \eqref{eq:ATk}, the dominant magnetization $m^*$
satisfy \eqref{mstarequation} and \eqref{eq:AT} for general
distribution of $H$.  This can be made precise by formulating a
variational problem.

We also note an alternative technique for evaluating the free energy
\eqref{eq:Zh} using Fourier transform and saddle point method, which
is standard in statistical mechanics (often without rigorous
justification).  Usage of this technique in information theory can be
found in e.g., \cite{Nishim01}.


\section*{Appendix B -- Estimation of Sparse Signals: An Alternative Derivation of \eqref{mutualinfo}}
\label{app:I}

In case of quadratic exponent \eqref{quadratic}, the partition
function \eqref{eq:Zbh} can be written using the Hubbard--Stratonovich
transformation as
\begin{align}
\label{gaussianintegral}
\sum_{\bmu} P(\bmu)e^{\sum_i\mu_ih_i}&=
C_n 
\sum_{\bmu}\exp\left\{a\sum_i\mu_i+\sum_i\mu_ih_i+\frac{b}{2n}
\bigg(\sum_i\mu_i\bigg)^2
\right\}\nonumber\\
&=C_n \sqrt{\frac{nb}{2\pi}} 
\int_{-\infty}^\infty \intd m
\exp\left\{-\frac{nbm^2}{2}\right\}
\sum_{\bmu}\exp\left\{a\sum_i\mu_i+\sum_i\mu_ih_i+
bm\sum_i\mu_i\right\}\nonumber\\
&=C_n \sqrt{\frac{nb}{2\pi}} 
\int_{-\infty}^\infty \intd m
\exp\left\{-\frac{nbm^2}{2}\right\}\prod_{i=1}^n\left[2\cosh\left(
a+bm+h_i\right)\right]\nonumber\\
&=C_n \sqrt{\frac{nb}{2\pi}}
\int_{-\infty}^\infty \intd m
\exp\left\{n\bigg[-\frac{bm^2}{2}+
\frac{1}{n}\sum_{i=1}^n\ln[2\cosh(a+bm
+h_i)]\bigg]\right\}.
\end{align}
Thus, we have $-\ln\hat{Z}\approx n\min_m \psi(m)-\ln C_n$, where
$\psi$ is defined by \eqref{eq:psistar},
whose minimum is attained at $m^*=m^*(\beta)$, one of the solutions to the
equation
$m=\bE\{\tanh(bm+a+H\}$,
as before.\footnote{The function $\psi(m)$ is (within a factor of the inverse
temperature) identified with
the Landau free energy function for this problem \cite[p.\ 186, eq.\ (4.15a)]{NO88},
\cite[Sect.\ 4.6]{Honerkamp02}.}
The mutual information is then obtained as \eqref{mutualinfo}.

\section*{Appendix C -- Estimation of Sparse Signals: The MMSE}
\label{app:E}

The MMSE is equal to twice the derivative of (\ref{mutualinfo}) 
w.r.t.\ $\beta$.
We will denote hereafter
$H_i$ as given by \eqref{eq:hi} with $y_i$ replaced by $Y_i$
and $\bH=(H_1,\ldots,H_n)$.
Let us present the asymptotic MMSE per sample,
$\lim_{n\to\infty}\mbox{mmse}(\bX|\bY)/n$,
as $A+B$, where $A$ is the double derivative of the first three terms,
and $B$ is the contribution of the other terms. The easy part is
the former:
$$A=\frac{\sigma^2q}{2(1+q)^2}+\frac{(1-m_a)\sigma^2}{2}\left[1-\frac{q(1+q/2)}
{(1+q)^2}\right].$$
As for $B$, we have the following consideration: The first three terms
depend only on $m_a$, which in turn is independent of $\beta$, therefore their
derivatives w.r.t.\ $\beta$ all vanish. For the last two terms, pertaining
to $\psi(m^*)$, it proves useful to return to the original expression
of the Gaussian integral (\ref{gaussianintegral}), i.e.,
\begin{align}
B&=-\frac{2}{n}\frac{\partial}{\partial\beta}\bE\{\ln \hat{Z}(\beta|\bH)\}
\nonumber\\
&=-\frac{2}{n}\frac{\partial}{\partial\beta}\bE\left\{\ln
\int_{-\infty}^\infty\frac{d\nu}{\sqrt{2\pi}}
\exp\left\{n\left[-\frac{(\nu-a)^2}{2b}+
\frac{1}{n}\sum_{i=1}^n\ln[2\cosh(
\nu+h_i)]\right]\right\}\right\}\nonumber\\
&=-\frac{2}{n}\frac{\partial}{\partial\beta}
\int_{\reals^n}d\by P_\beta(\by)\ln
\int_{-\infty}^\infty \intd m
\exp\left\{n\left[-\frac{bm^2}{2}+
\frac{1}{n}\sum_{i=1}^n\ln[2\cosh(bm+a
+h_i)]\right]\right\}\nonumber\\
&=-\frac{2}{n}
\int_{\reals^n}d\by\frac{\partial P_\beta(\by)}{\partial\beta}\ln
\int_{-\infty}^\infty \intd m
\exp\left\{n\left[-\frac{bm^2}{2}+
\frac{1}{n}\sum_{i=1}^n\ln[2\cosh(bm+a
+h_i)]\right]\right\}\nonumber\\
& \qquad -
\frac{2}{n}\int_{\reals^n}d\by P_\beta(\by)\frac{\partial}{\partial\beta}\ln
\int_{-\infty}^\infty \intd m
\exp\left\{n\left[-\frac{bm^2}{2}+
\frac{1}{n}\sum_{i=1}^n\ln[2\cosh(bm+a
+h_i)]\right]\right\}\nonumber\\
&\eqd B_1+B_2.
\end{align}
Now, $P_\beta(\by)$ is the mixture of Gaussians weighted by 
$\{P(\bmu)\}\}$, where the dominant $\bmu$--configurations are
those with $(1+m_a)/2$ $(+1)$'s and
$(1-m_a)/2$ $(-1)$'s. Each such configuration contributes the same
quantity to $B_1$ and $B_2$, because for every given such $\bmu$,
the random variables $\{Y_i\}$ (and hence also $\{H_i\}$) are all independent,
a fraction $(1+m_a)/2$ of them are $\calN(0,1/\beta)$ and
the remaining fraction of $(1-m_a)/2$ are $\calN(0,\sigma^2+1/\beta)$.
Thus, it is sufficient to confine 
attention to one such sequence, call it $\bmu^*$, whose
first $n_1\eqd n(1-m_a)/2$ components are all $-1$ and last
$n-n_1= n(1+m_a)/2$ components are all $+1$.
Thus,
\begin{align}
B_1&\approx -\frac{2}{n}
\int_{\reals^n}d\by\frac{\partial P_\beta(\by|\bmu^*)}{\partial\beta}\ln
\int_{-\infty}^\infty \intd m
\exp\left\{n\left[-\frac{bm^2}{2}+
\frac{1}{n}\sum_{i=1}^n\ln[2\cosh(bm+a
+h_i)]\right]\right\}\nonumber\\
&\approx \frac{1}{n}
\mbox{Cov}\left\{\sum_{i=1}^{n_1}Y_i^2
+\frac{1}{(1+q)^2}\sum_{i=n_1+1}^nY_i^2,
\sum_{i=1}^n\ln[2\cosh(bm^*+a
+H_i)]\right\}\nonumber\\
&=\frac{1+m_a}{2}\cdot\mbox{Cov}_0\{Y^2,\ln[2\cosh(bm^*+a+H)]\}\nonumber\\
& \qquad + \frac{1-m_a}{2}\cdot\frac{1}{(1+q)^2}\cdot
\mbox{Cov}_1\{Y^2,\ln[2\cosh(bm^*+a+H)]\}.
\end{align}
where $\mbox{Cov}_s\{\cdot,\cdot\}$ denotes covariance with respect to
$\calN(0,\sigma^2s+1/\beta)$, $s=0,1$.
Finally, for $B_2$, we have:
\begin{align}
B_2&=-\frac{2}{n}\int_{\reals^n}d\by P_\beta(\by)
\frac{\partial}{\partial\beta}\ln
\int_{-\infty}^\infty \intd m
\exp\left\{n\left[-\frac{bm^2}{2}+
\frac{1}{n}\sum_{i=1}^n\ln[2\cosh(bm+a
+h_i)]\right]\right\}\nonumber\\
&=\frac{1}{n}\int_{\reals^n}d\by P_\beta(\by)
\cdot\frac{\int_{-\infty}^\infty \intd m\left[\sum_ih_i'\tanh(bm+a+h_i)\right]
e^{-n\psi(m)}}{
\int_{-\infty}^\infty \intd m e^{-n\psi(m)}}\nonumber\\
&\approx
\bE\left\{\frac{1}{n}\sum_{i=1}^n H_i'\tanh(bm^*+a+H_i)\right\}\nonumber\\
&\approx\frac{1+m_a}{2}\cdot\bE_0\{H'\tanh(bm^*+a+H)\}+
\frac{1-m_a}{2}\cdot \bE_1\{H'\tanh(bm^*+a+H)\},
\end{align}
where $\bE_s$ denotes expectation w.r.t.\
$\calN(0,\sigma^2s+1/\beta)$, $s=0,1$, and $H'$ is given by
\eqref{eq:hip}, and correspondingly, $h_i'$ 
and $H'_i$ are given by the same formula with
$Y$ replaced by $y_i$ and $Y'_i$ respectively.
Collecting all terms, $A$, $B_1$, and $B_2$, we 
have \eqref{eq:mmse}.

\section*{Appendix D -- Estimation of Sparse Signals: Two Extreme Cases}
\label{app:ex}

Two extreme cases, where it is
relatively easy to examine the resulting expression are as follows:
\begin{itemize}
\item When
 $b \gg 1$ and $a \ll -1$, we have $m_a \approx  -1$ and
$m^* \approx -1$ (which means that most $s_i=1$), and so we
can approximate
$$\ln[2\cosh(bm^*+a+H)]\approx \ln[2\cosh(-b+a+H)]
\approx b-a-H$$ 
and $\tanh(bm^*+a+H)\approx -1$,
and we get
$$\lim_{n\to\infty}\frac{\mbox{MMSE}(\bX|\bY)}{n}\approx \frac{\sigma^2}{1+q},$$
the classical Wiener expression, as expected.\footnote{Here, by
$\lim_{n\to\infty}\mbox{MMSE}(\bX|\bY)/n\approx F(a,b,\beta,\sigma^2),$ for a generic function
$F$, we mean that $\lim_{a\to-\infty}\lim_{b\to\infty}\lim_{n\to\infty}
nF(a,b,\beta,\sigma^2)/\mbox{MMSE}(\bX|\bY)=1$. A similar comment applies to
item number 2 below.}
\item When $b \gg 1$ and $a \gg 1$, we have $m_a \approx  1$ and
$m^* \approx 1$ (which means that most $s_i=0$), and then 
$\ln[2\cosh(bm^*+a+H)]\approx b+a+H$ and 
$\tanh(bm^*+a+H)\approx 1$, so we get
$$\lim_{n\to\infty}\frac{\mbox{MMSE}(\bX|\bY)}{n}\approx \frac{1-m_a}{2}\cdot\sigma^2,$$
which means the conditional--mean estimator simply outputs essentially the all--zero sequence
without attempting to detect (explicitly or implicitly) which of the few signal
components are active. The intuition behind this behavior is that
when there are so few active components of the clean signal,
then even if there are nevertheless a few observations
$\{y_i\}$ with large absolute values (and hence could have been suspected
to stem from places where $s_i=1$),
it is still more plausible for the estimator to ``assume'' that they
simply belong to the tail of $\calN(0,1/\beta)$ (with $s_i=0$) rather than
to $\calN(0,\sigma^2+1/\beta)$ with $s_i=1$. This because the prior for $s_i=1$
is so small that it becomes comparable to the tail probability of $\calN(0,1/\beta)$.\footnote{
To see this, it is instructive to think of a simple binary hypothesis
testing problem where an observer is required to decide whether an observation
comes from $\calN(0,1/\beta)$ or $\calN(0,\sigma^2+1/\beta)$ and the priors are
very much in favor of the former.}
\end{itemize}

\end{document}